\documentclass[a4paper,times]{emsart}

\usepackage{amsthm} 

\usepackage{amsmath,amssymb,amsfonts}
\usepackage{graphicx}
\usepackage{textcomp}
\usepackage{subfigure}
\usepackage{algorithm,algorithmic}
\usepackage{multirow}

\journalname{Arxiv}
\startpage{1}

\title[Reconfigurable Metasurface: A Systematic Categorization and Recent Advances]{Reconfigurable Metasurface: A Systematic Categorization and Recent Advances}

\author{%
Changhao Liu\affilnums{1,2}, 
Fan Yang\affilnums{1,2}, 
Shenheng Xu\affilnums{1,2}, and
Maokun Li\affilnums{1,2}
}

\affiliation{%
\affilnum{1}Department of Electronic Engineering, Tsinghua University, Beijing, 100084, China.\\
\affilnum{2}Beijing National Research Center for Information Science and Technology (BNRist), Beijing, 100084, China.
}

\corrauth{Fan Yang}
\email{fan\_yang@tsinghua.edu.cn}

\receivetime{Received March 22, 2022}
\accepttime{Accepted March 22, 2022}
\publishtime{Published March 22, 2022}

\authorcopy{C. Liu \emph{et al.}}

\abstract{%
Considering the rapid progress of theory, design, fabrication and applications, metasurface (MTS) has become a new research frontier in microwave, terahertz and optical bands. Reconfigurable metasurface (R-MTS) can dynamically modulate electromagnetic (EM) wave with unparalleled flexibility, which leads to great research tide in recent years. Numerous R-MTSs with powerful capabilities and various functions are presented explosively. In light of the five dimensions of EM wave, this review proposes a unified model to describe the interactions among R-MTS, EM wave and EM information, and suggests information bit allocation strategy to categorize different types of R-MTSs systematically. As recent advances of R-MTS, 1-bit and 2-bit elements manipulating different wave dimensions are reviewed respectively in detail. Finally, this review discusses the future research trends of R-MTS. Hopefully, R-MTSs with diverse dimensions and functions can propel the next generation of communication, detection, sensing, imaging and computing applications.
}


\begin{document}

\maketitle

\section{Introduction}
Electromagnetic (EM) wave is the cornerstone of modern society. To manipulate EM wave, metasurface (MTS), an artificial array composed of ultra-thin subwavelength elements, is invented and investigated over the decades. MTS can manipulate multi-dimensional EM wave with unparalleled degree of freedom, which has great advantages compared with conventional bulk device manipulating EM wave, like low cost, low profile, light weight, high efficiency and easy fabrication. Therefore, MTS is attracting tremendous interest of researchers nowadays, and applications around MTS are experiencing explosive growth, such as high performance antennas \cite{reflectarray}, radar cross section (RCS) reducing radomes \cite{RCS1}, invisibility cloaks \cite{cloak1}, holograms \cite{hologram1, hologram2}, ultra-thin lenses \cite{metalens}, to name a few.

Evolving from conventional MTS, reconfigurable metasurface (R-MTS), integrated with tunable materials on each element, shows dynamic capability to control EM wave. Scattered patterns of EM wave after encountering R-MTS can be tuned by applying input stimuli to R-MTS. A variety of R-MTS are proposed with the reconfigurable functions of tuning different dimensions of EM wave. Moreover, some state-of-the-art prototypes are emerging with multiplexed functions on one R-MTS, which shows that the research of R-MTS are evolving toward multi-dimensions and multi-functions.

Digital R-MTS, a reconfigurable form of coding metasurface \cite{1bitcodingmeta}, brings about huge research tide. Loading lumped switches with discrete controlling signals on MTS shows great advantages over conventional analog R-MTS, due to its low cost, easy manipulation, scalability, high reliability, and compatibility with information world. Therefore, this review mainly focuses on digital R-MTS, which is characterized by information bit.

Digital R-MTS not only shapes the physical EM wave and guides the EM energy flow, but also modulates the incident wave and generates information. Information is loaded on EM wave, and has various forms, depending on the various dimensions of EM wave. Consequently, R-MTS is given more functions and capabilities to manipulate multi-dimensions of EM wave, which also brings the great boom of R-MTS investigation. 

With tremendous advantages and revolutionary applications, numerous R-MTSs with diverse functions are just unfolding. However, various terminologies based on their functions appear correspondingly, which make the architecture of R-MTS research seem complex and confusing. Therefore, a systematic categorization of R-MTSs is in need. Here, a unified mathematical model of R-MTSs is propose based on five dimensions of EM wave to describe the interactions among R-MTS, EM wave and EM information, and then we suggest a method called \textit{information allocation strategy}, attempting to reorganize these R-MTSs under one terminology system. Based on this systematic catergorization, recent advances of R-MTSs are reviewed within the same terminology system. This strategy could contribute to the architecture of surface electromagnetics theory \cite{EMsurface}. Hopefully, this review could provide a comprehensive view of R-MTS, EM wave, EM information and their interactions, and inspire researchers to discover more fascinating functions of R-MTS. 

This review is organized as follows. Firstly, we focus on the interactions among EM wave, EM information and R-MTS, and propose an information allocation strategy in Section \ref{secIII}. Next, under the information allocating frame, different forms of 1-bit allocation elements are reviewed in Section \ref{secIV}. Section \ref{secV} continues the point of view, and it mainly reviews the 2-bit allocation elements. In Section \ref{secVI}, the future trends of R-MTS are discussed, and some challenges are also outlined. Conclusion is drawn in Section \ref{secVII} as the final part.

\section{Information Allocation Strategy}
\label{secIII}

\begin{figure*}[!t]
	\centerline{\includegraphics[width=1.5\columnwidth]{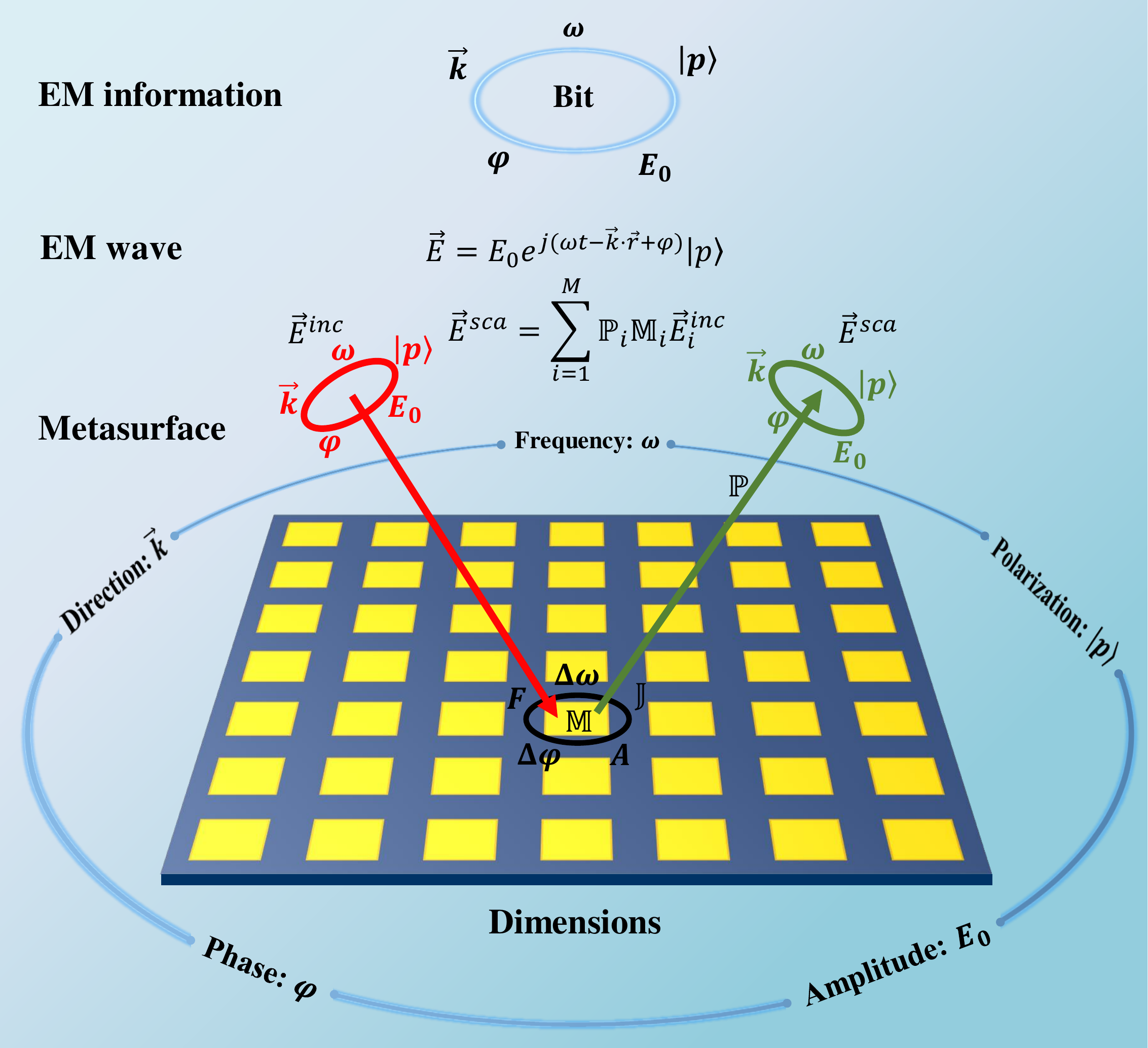}}
	\caption{An illustration of information allocation strategy and the interactions among EM information, EM wave and R-MTS, which are linked by EM dimensions. Since EM wave has five dimensions, EM information bits can be allocated to these dimensions. When an incident EM wave containing five dimensions encounters an R-MTS, it is modulated by a modulation matrix $\mathbb{M}$ of each element, which also manipulates five EM dimensions and loads EM information on EM wave. After propagating in free space, the total scattered wave in far field is a summation of all scattered waves from each element. In summary, information to be transmitted is allocated to dimensions of scattered EM wave, and modulated dynamically by R-MTS. }
	\label{bitallocation1}
\end{figure*}

\subsection{Dimensions of EM Wave}

To begin with, it is necessary to review the dimensions of plane EM wave. A plane EM wave can be mathematically described in the following formula
\begin{equation}
	\vec{E}(t,\vec{r})=E_{0}e^{j(\omega t-\vec{k} \cdot \vec{r} + \varphi)} |p\rangle.
	\label{planewave}
\end{equation}
This formula contains five dimensions of plane EM wave: phase ($\varphi$), amplitude ($E_{0}$), polarization ($|p\rangle$), direction ($\vec{k}$) and frequency ($\omega$). Note that field has polarization, so $\vec{E}$ is a vector, also as a function of space ($\vec{r}$) and time ($t$).

\subsection{Interactions among R-MTS, Wave and Information}

To understand the functions of R-MTS, the physical point of view is interpreted at first. When EM wave encounters R-MTS, each element on R-MTS scatters local EM wave. Then the scattered EM waves propagate in free space. Summing up every local scattered EM wave, the receiver obtains the final EM response. R-MTS can manipulate EM wave dynamically in five dimensions. Accordingly, R-MTS enables the dynamic control of wave-metasurface interactions.

If information viewpoint is taken into consideration, another understanding of the functions of R-MTS can be obtained. We mainly discuss digital information in this review. Digital information is fundamental in modern world, which is characterized by bit. Compared with analog information, digital information is robust, fast, low-cost and easy to process, which is more and more important nowadays. 
To generate and transmit digital EM information in free space, the physical information medium, EM wave, is indispensable. In other words, EM wave carries EM information. Since EM wave has five dimensions, EM information has five dimensions correspondingly. Besides, each dimension of EM wave can carry EM information respectively. R-MTS is invented to manipulate EM wave, so using R-MTS to manipulate EM wave is to modulate EM information. Since EM information and EM wave have five dimensions, R-MTS also has to support the function of modulating five dimensions of EM information. Therefore, such type of R-MTS is multi-dimensional. Figure. \ref{bitallocation1} illustrates interactions among R-MTS, EM wave and EM information, linked by EM dimensions. 

EM elements compose R-MTS, and tunable material enables the reconfiguration of EM element. Consequently, the key to modulating digital EM information lies in designing EM elements with tunable materials. Since the information to be generated is discrete and digital, the tuning material is specified as digital switch, which generates discrete states of elements, and is also characterized by bit in physical world correspondingly. Compared with general tunable materials, digital switches are cost-effective, easy to integrate, easy to control and compatible with digital information. Therefore, R-MTSs using digital switches attract enormous research interest in recent years.

Based on the information point of view, the R-MTS provides dynamic control of information-metasurface interactions. R-MTS can not only manipulate physical EM wave, but can also generate and modulate EM information. R-MTS links physical world and information world, which is named as \textit{information metasurface} \cite{informationmeta}.

\subsection{Mathematical Model of R-MTS}

Here, we propose a mathematical model of R-MTS to describe the interactions among R-MTS, EM wave and EM information. Suppose an R-MTS is composed of $M$ elements. When an EM wave encounters R-MTS, the scattered wave is expressed as the following formula
\begin{equation}
	\label{response}
	\vec{E}^{sca}(t, \vec{r})=\sum_{i=1}^{M} \mathbb{P}_i(\vec{r}) \cdot \mathbb{M}_i(t, \vec{r}, R_i) \cdot \vec{E}^{inc}_i(t, \vec{r}_i),
\end{equation}
where $inc$ represents the incident wave, and $sca$ means the scattered wave. $\vec{r}$ is the location of the receiver, and $\vec{r}_i$ is the position of the $i$th element. Here, $\vec{E}^{inc}_i$ is the incident wave at $i$th element. Note that the incident wave is not necessarily plane wave, but it is assumed at small local region of each element, the incident wave is plane wave. According to (\ref{planewave}), $\vec{E}^{inc}_i$ can be expressed as
\begin{equation}
	\vec{E}^{inc}_i(t, \vec{r}_i) = E_{0i}^{inc}e^{j(\omega_i^{inc} t-\vec{k}^{inc}_i \cdot \vec{r}_i + \varphi_i^{inc})} |p_i^{inc}\rangle,
\end{equation}
with variable values of incident dimensions at each element of different positions.

$\mathbb{M}_i(t, \vec{r}, R_i)$ is the $i$th element modulation matrix of the incident wave, and the notation
\begin{equation}
	\label{Ri}
	R_i=(\vec{E}^{inc}_i, \rm{Switch}_i)
\end{equation}
 is used to represent the response of elements. $R_i$ contains the incident dimensions of EM wave and the element configurations determined by switch. We call the former \textit{excitation} and the latter \textit{control}. It is also remarked that $R_i$ is also a funtion of time ($t$), because $\rm{Switch}_i$ can be modulated temporally, which is actually the origin of reconfiguration. 

$\mathbb{M}_i(t, \vec{r}, R_i)$ is expressed as
\begin{equation}
	\label{elementmodulation}
	\mathbb{M}_i(t, \vec{r}, R_i)=A_i(R_i) F_i(\vec{r}, R_i) e^{j(\Delta\omega_i(R_i) t + \Delta\varphi_i(R_i))} \mathbb{J}_i(R_i),
\end{equation}
where the amplitude response ($A$) describes the energy amplifying, attenuating or maintaining states; element radiation pattern ($F$) is a function of the receiving location ($\vec{r}$), and can shape the beam as well as determine the beam direction; additional frequency ($\Delta\omega$) can shift the incident frequency, which is a kind of nonlinear effect; additional phase ($\Delta\varphi$) describes the phase shifting effect when the incident wave encounters EM element; Jones polarization matrix ($\mathbb{J}$) is a $2\times2$ matrix, usually anisotropic, which can modulate the incident polarization ($|p^{inc}\rangle$), and $\mathbb{M}$ is also a $2\times2$ matrix accordingly. These five element modulation dimensions can exactly cover and modulate the five dimensions of EM wave.

$\mathbb{P}_i(\vec{r})$ describes the path from R-MTS to the information receiver. Suppose the path is linear, time-invariant, isotropic, homogeneous and single-path, so the mathematical model is
\begin{equation}
	\mathbb{P}_i(\vec{r}) = L_i(\vec{r}) e^{-j\vec{k} \cdot (\vec{r}-\vec{r}_i)}\mathbb{I},
\end{equation}
where $L_i$ is the path loss from the $i$th element to the receiver, and $\mathbb{I}$ is the polarization identity matrix. Obviously, the path model is determined by the location of information receiver ($\vec{r}$) and the position of each element ($\vec{r}_i$).

Each element may have different responses to different dimensions of incident wave, and since the elements can be reconfigurable in several dimensions, these dimensions of EM wave can be dynamically modulated. Summing up each response of element, the EM wave scattered by R-MTS is obtained according to (\ref{response}).

\subsection{Information Allocation Strategy}

As shown in (\ref{response}), the characteristics of R-MTS are determined by EM elements. Suppose N-bit information ($2^N$ states) needs to be generated by an EM element. Since EM information rests in EM wave, the N-bit information can be allocated to different dimensions of EM wave. Correspondingly, those dimensions of EM wave need to be independently modulated by the element and each element has to respond to the desired dimensions of EM wave. Meanwhile, loading $N$ independent switches enables N-bit element. The key to transmitting multi-dimensional EM information is designing multi-dimensional reconfigurable elements correspondingly. 
This is why this strategy is named as \textit{information bit allocation}, which means allocating multi-dimensional EM information bits to multi-dimensional reconfigurable EM elements. This strategy can help researchers to understand the functions of each type of reconfigurable element, and find the category of various reconfigurable elements.

Strictly speaking, the total EM response is the summation of each element of different dimensions. Since there are $M$ independent elements in an R-MTS, and each element manipulates $N$ independent bits, the total modulated bits of the entire R-MTS are $M\times N=K$, which can greatly multiply the amount of information. Considering the effect of array, the arrangements and states of each element at array level also worth careful design, which is called array-level bit allocation. But we mainly focus on the element level in this review due to its importance and fundamentality. The array-level bit allocation strategy is more complex and flexible, which is beyond the scope of this review.

In the next sections, the information allocation strategy is applied to review the reconfigurable elements of different dimensions. We first review 1-bit elements which have only one reconfigurable dimension, and it is usually enabled by one independent switch. Then the 2-bit elements are reviewed, which may have two reconfigurable independent dimensions enabled by two independent switches. Following this path, more-bit reconfigurable elements of different dimensions can be designed to realize more functions.

\section{1-Bit Element}
\label{secIV}
In this section, 1-bit reconfigurable elements which have only one independent switch are reviewed. Limited by one bit, each element has two states. So 1-bit element only manipulates one dimension of EM wave, or several associated dimensions but with only 1-bit information or two states totally.

\begin{figure}[!t]
	\centering
	\subfigure[] { 
		\includegraphics[width=0.45\columnwidth]{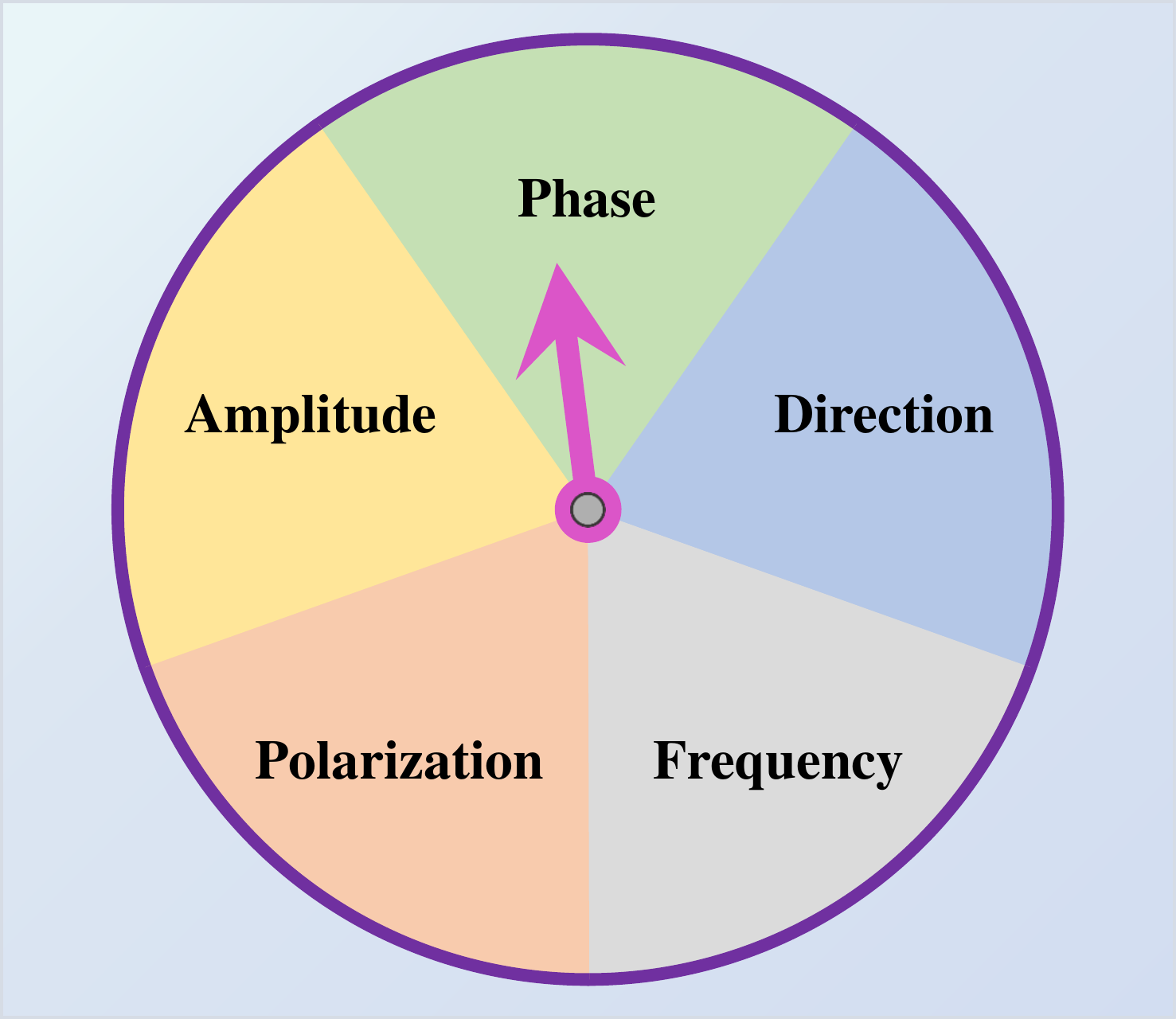} 
	} 
	\subfigure[] {
		\includegraphics[width=0.45\columnwidth]{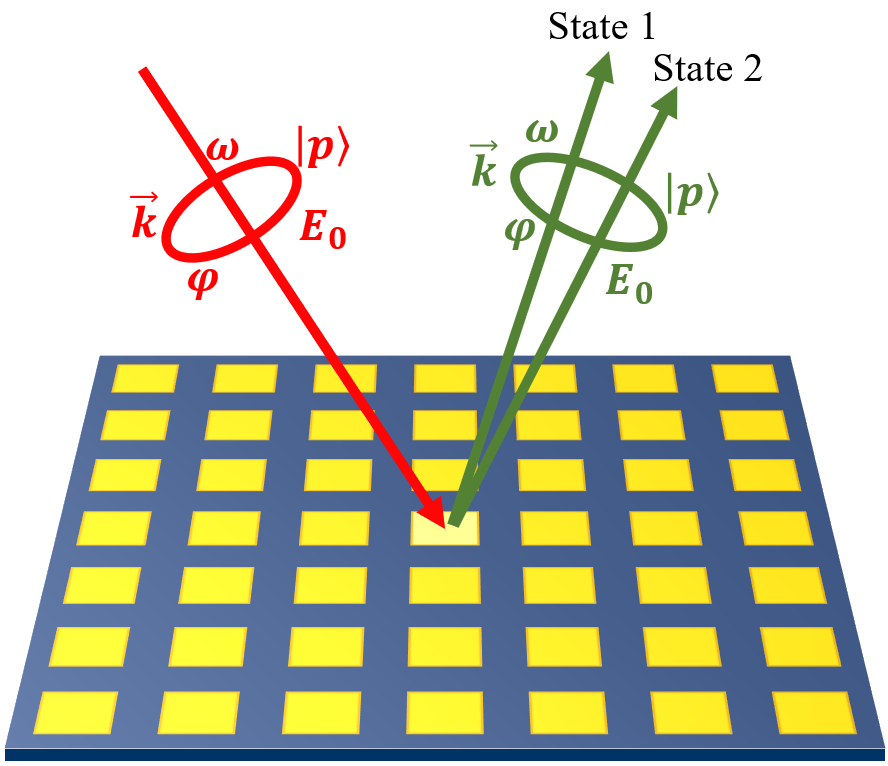} 
	} 
	\caption{An illustration of 1-bit allocation. (a) 1 bit information can be allocated to one of five dimensions at a time. So there are five types of 1-bit elements. (b) The function of the 1-bit reconfigurable elements, where the element can manipulate one incident wave with two different scattering states.}
	\label{1bit}
\end{figure}

Since there are five dimensions of EM wave, there are five types of information bit allocation, specified as phase-only, amplitude-only, polarization-only, direction-only and frequency-only elements (Figure. \ref{1bit}). 
Researchers have devoted significant efforts to design each type of reconfigurable elements, and have achieved major breakthrough in the past decades. Some typical achievements of their works are reviewed in the following parts respectively.

\begin{figure*}[!t]
	\centerline{\includegraphics[width=1.8\columnwidth]{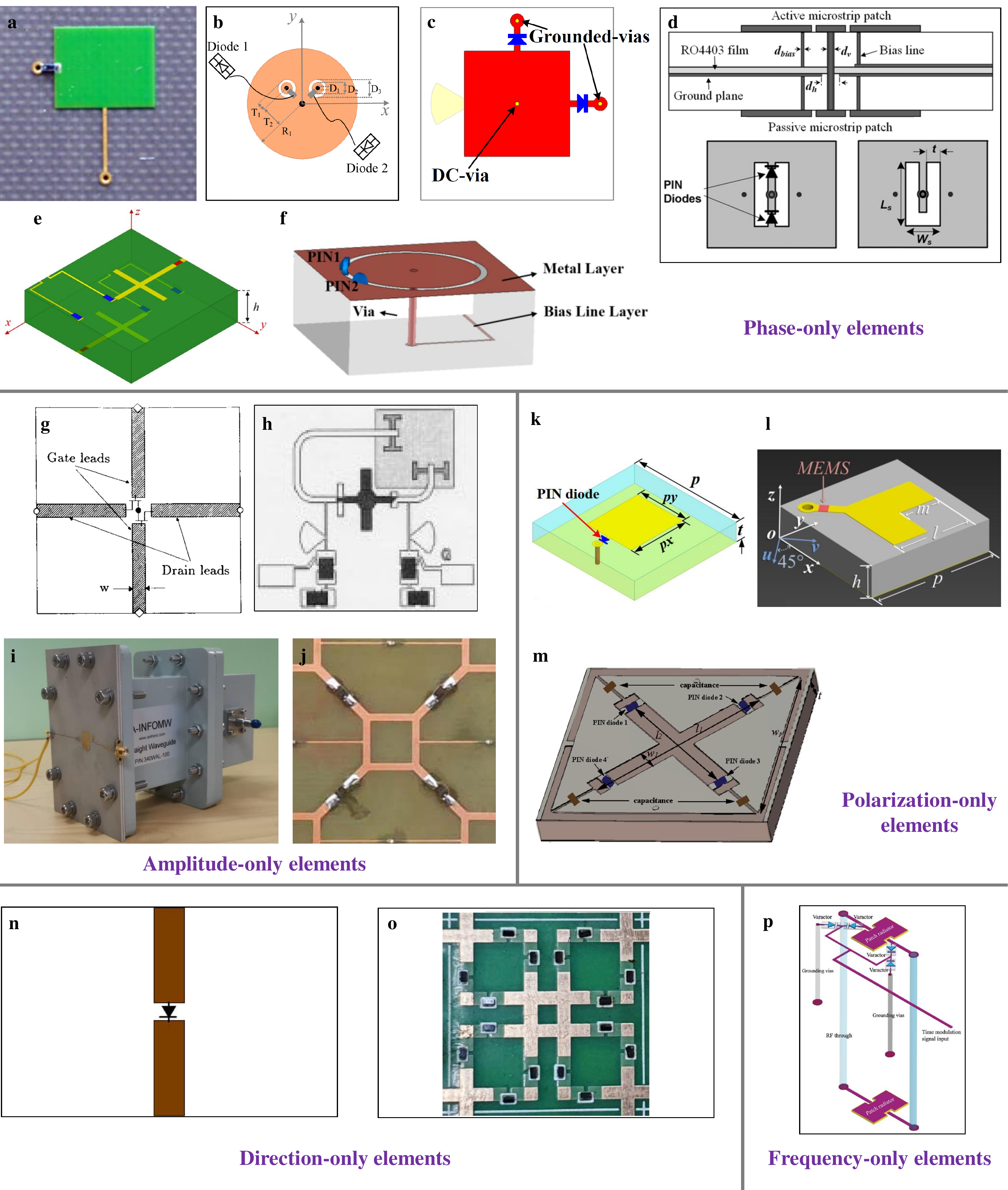}}
	\caption{1-bit reconfigurable elements manipulating one of five dimensions. 
		(a)-(f) Phase-only elements. 
		(a) 1-bit phase reconfiguration with single switch on each element [17]. 
		(b) 1-bit phase reconfigurable reflective element with two switches. It can realize polarization conversion [23]. 
		(c) 1-bit element with two switches controlled by one bias signal can respond to dual LPs and CPs [28]. 
		(d) Reconfigurable transmitarray element based on receive-transmit structure. Two PIN diodes are turned ON/OFF alternately, leading to 180$^\circ$ current reversal on transmitting structure [30]. 
		(e) A dual-layer reconfigurable transmitarray element with one switch on one layer to tune the phase simultaneously [39]. 
		(f) Single layer bidirectional reconfigurable element [40]. 
		(g)-(j) Amplitude-only elements. 
		(g) Amplifying grid array based on differential amplifier [45]. 
		(h) Amplifying reflectarray based on FET transistors [46]. 
		(i) An amplifying element using the principle of parametric amplification [47]. 
		(j) A switchable absorber with four PIN diodes on different directions has the ability to respond to full polarizations [52]. 
		(k)-(m) Polarization-only elements. 
		(k) LP-LP switching element [10]. 
		(l) LP-CP switching element [59]. 
		(m) CP-CP switching element [60]. 
		(n)-(o) Direction-only elements. 
		(n) Cylindrical active dipole elements on an FSS [61]. 
		(o) Complementary \#-shaped 1-bit reconfigurable planar FSS in full polarization [64]. 
		(p) Frequency-conversion element based on time-modulation [70]. 
	}
	\label{1bitsummary}
\end{figure*}

\subsection{Phase-Only Element}

Phase-only elements are most intensively studied type among the five types of 1-bit R-MTS. Considering phase-only elements, the formula (\ref{response}) is simplified into the following form
\begin{equation}
	\label{phase-only}
	\vec{E}^{sca}(\vec{r},t)=\sum_{i=1}^{M} \vec{C}_i(\vec{r}) e^{j(-\vec{k} \cdot (\vec{r}-\vec{r_i}) + \Delta\varphi_i(R_i)+\varphi_i^{inc})}\cdot e^{j\omega t},
\end{equation}
with time-invariant coefficient term
\begin{equation}
	\vec{C}_i(\vec{r})=L_i(\vec{r}) A_i  E_{0i}^{inc} F_i(\vec{r})\mathbb{J}_i |p_i^{inc}\rangle.
		\label{phase-only-invar}
\end{equation}
Phase-only elements only change the $i$th additional phase $\Delta\varphi_i$, and other dimensions are usually invariant with switch configurations and time. Formula (\ref{phase-only}) indicates that the radiation pattern of R-MTS is determined by $\Delta\varphi$ of each element, which means the near field phase influencies far field beam patterns. Therefore, various applications are realized by phase-only R-MTSs, such as dynamic beamforming and beam scanning \cite{1bitphaseorpol}, reconfigurable hologram \cite{1bitphasehologram}, reconfigurable orbit angular momentum (OAM) beam generation \cite{1bitOAM}, and so on.

1-bit phase is the quantization of continuous phase, which has only two phase states. Usually, researchers choose the phase difference of two configurations of near 180$^\circ$, because the phase quantization loss can be minimized to about 3 dB - 3.9 dB \cite{phasequan,phasequan2}.

Various 1-bit elements are designed, with different configurations, switches, frequency bands and functions. According to reflectarray theory, additional phase of elements can be tuned by time-delay lines, variable sizes or variable rotating angles \cite{reflectarray,summary2,summary1}. In light of these phase tuning approaches, ON/OFF states of RF switch on element can be used to mimic the size changing or angle rotating. Therefore, the additional phase of elements can be changed.

\subsubsection{Basic Phase-Only Element}
One switch with 1-bit controlling signal can enable phase-only reconfiguration for single polarization EM wave. 
In microwave band, single PIN diode is often used as the switch to tune the phase \cite{1bitcodingmeta,1bitphase9,1bitphase11mulfreq,1bitphaseorpol,1bitphase1860G,1bitphase12W,1bitphase15big,1bitphasewideband}. In \cite{1bitphase9,1bitphase11mulfreq,1bitphaseorpol}, independently addressable reconfigurable reflectarray antenna (RRA) is realized with high efficiency by carefully designing the elements, switches and bias lines (Figure. \ref{1bitsummary}(a)). Besides, various RRAs are brought out with different bands. In millimeter wave band, a 60 GHz RRA prototype using commercial PIN diodes is fabricated \cite{1bitphase1860G}. Wideband RRAs are always a research focus, and up to 38.4\% bandwidth is reported in a recent work \cite{1bitphasewideband}.
In THz and optical bands, the modulation principles are similar. However, the immaturity of tunable devices limits the development in this research field. In recent years, numerous tunable materials are also introduced to tune the phase at such high frequency bands. 

Manipulating several switches synchronously using one controlling signal with two states is also considered as a 1-bit element. More functions can be integrated onto one element with more associated switches. 

\subsubsection{Phase-Only Element Related With Polarization Dimension}

Polarization conversion with stable 180$^\circ$ phase modulation in full band can be realized by loading more than one associated switch. Jones matrix with polarization conversion function is derived if elements and switches are designed properly with
\begin{equation}
	\label{jones}
	\mathbb{J} = \begin{bmatrix}
		0 & 1 \\ 1 & 0
	\end{bmatrix},
\end{equation}
which means this Jones matrix can swap the incident polarization, and in other words, the reflection polarization is rotated by 90$^\circ$. Applying controlling signal to alternately tune the ON/OFF states of switches can realize accurate 180$^\circ$ phase difference. Lots of works have been done using the principle of polarization conversion in RRA designs \cite{1bitphase2,1bitphase3,1bitphase5convpol,1bitphase1,1bitphase8convpol4PIN}. For instance, literature \cite{1bitphase2} demonstrates that alternately switching two PIN diodes controlled by one signal line can realize stable 180$^\circ$ reflection phase difference based on the principle of polarization conversion (Figure. \ref{1bitsummary}(b)), which can manipulate circular polarization (CP) wave with 1-bit phase difference. 

Dual-switch elements can respond to dual LP and dual CP waves. Polarization conversion elements with two or more switches can respond to multi-polarizations. Besides, literature \cite{1bitphase10mulpol} reports that a patch loading two associated switches on the orthogonal sides can tune phase with response to dual-linear and dual-circular polarizations (Figure. \ref{1bitsummary}(c)).

\subsubsection{Phase-Only Element Related With Direction Dimension}

Apart from reflective metasurfaces, transmissive metasurfaces with 1-bit phase tuning ability are also investigated comprehensively, which are called reconfigurable transmitarray antenna (RTA). It has been proved that at least two switches are necessary to realize a high-performance 1-bit RTA element \cite{1bitphaseRTAtheory}. Based on two switches, using current reversal method to realize 1-bit phase control is a useful approach when designing RTAs \cite{1bitphase4,1bitphase41,1bitphase14trans,1bitphase6trans,1bitphasetranscouple,1bitphase7convpoltrans,1bitphase7convpoltrans2}. For example, literature \cite{1bitphase4} proposes the receive-transmit structure with two PIN diodes changing the element configuration, which can reverse current by 180$^\circ$ mutually, and then the transmitted phases are reversed by 180$^\circ$ (Figure. \ref{1bitsummary}(d)). Unlike RTAs with all switches on a single layer, multi-layer RTAs with switches on different layers are proposed in recent years \cite{1bitphaseRTAtwolayer1, 1bitphaseRTAtwolayer2, 1bitphaseRTAtwolayer3}. For instance, a dual-layer Huygens' element is designed in literature \cite{1bitphaseRTAtwolayer3}, with two associated switches loading on each layer respectively to realize 1-bit phase control (Figure. \ref{1bitsummary}(e)).

Single layer R-MTS can also implement bidirectional beam scanning functions with polarization conversion \cite{1bitphase13transref}, where half of the energy is reflected and the other half is transmitted (Figure. \ref{1bitsummary}(f)).

\subsection{Amplitude-Only Element}

According to formula (\ref{elementmodulation}), amplitude modulation of element can be represented by $A$. When an EM wave encounters MTS, energy can be amplified ($A>1$), attenuated ($A<1$) or maintained ($A\approx1$), so the amplitude-only element should be active, lossy or near transparent. Amplitude modulation element can apply for channel modulation \cite{channelcodeamp}, relay amplifying \cite{relayamplify} and so on. Some microwave band 1-bit amplitude R-MTS prototypes are demonstrated in \cite{relayamplify,1bitampreview,1bitampampmaintain,1bitamp2ampmaintain,1bitamp3ampmaintainref,nonlinearthesis,1bitsummaryAFSS,1bitamp3rcs,1bitamp6LP,1bitamp5dualpol,1bitamp4allpol}.

Amplifying transmitarrays based on receive-transmit structure with active transistors as energy amplifier are proposed in 1990s \cite{1bitampreview,1bitampampmaintain,1bitamp2ampmaintain}. As Figure. \ref{1bitsummary}(g) illustrates, a pair of vertical gate leads are employed to receive the energy from free space, and two transistors compose a differential amplifier with their sources connected \cite{1bitamp2ampmaintain}. The gate energy is amplified and radiated horizontally by a pair of drain leads. Meanwhile, amplifying/maintaining states can be switched by tuning the ON/OFF states of power source of transistors.

Amplifying reflectarray based on FET transistors is proposed in \cite{1bitamp3ampmaintainref}, as shown in Figure. \ref{1bitsummary}(h). The patch receives LP wave and couples the energy into microstrip line through H-shaped slot. Then the energy is amplified in the circuit and re-radiates into free space through the patch in orthogonal polarization.

Recently, amplifying reflectarray based on parametric amplifier is investigated in \cite{relayamplify, nonlinearthesis}. 2.36 GHz incident signal is amplified and reflected based on the nonlinear effect of varactor on circuit with energy provided by 4.72 GHz pump (Figure. \ref{1bitsummary}(i)). The amplifying gain can be tuned by changing the pump energy, which enables the amplitude reconfiguration.

Switchable absorber-reflectors or absorber-transmitters based on active frequency selective surface (AFSS) have been developed in recent years \cite{1bitsummaryAFSS,1bitamp3rcs,1bitamp6LP,1bitamp5dualpol,1bitamp4allpol,1bitamp7allpol}. Imperfect PIN diodes with large insertion loss at ON state can absorb energy at resonance frequency, while at OFF state, the wave is scattered by the R-MTS without attenuation. Switching the ON/OFF states can modulate the amplitude. Elements can respond to multiple polarizations with several PIN diodes on them, from LP \cite{1bitamp3rcs,1bitamp6LP}, dual-LP \cite{1bitamp5dualpol}, to full polarizations (Figure. \ref{1bitsummary}(j)) \cite{1bitamp4allpol,1bitamp7allpol}.

\subsection{Polarization-Only Element}

Polarization $|p\rangle$ is one of the intrinsic dimensions of spatial EM wave. 1-bit polarization reconfigurable element switches between two polarization states, and mathematically, Jones matrix $\mathbb{J}$ changes between two forms. Polarization bits are also exploited to transmit information \cite{bitpol}. 

\begin{figure}[!t]
	\centerline{\includegraphics[width=0.5\columnwidth]{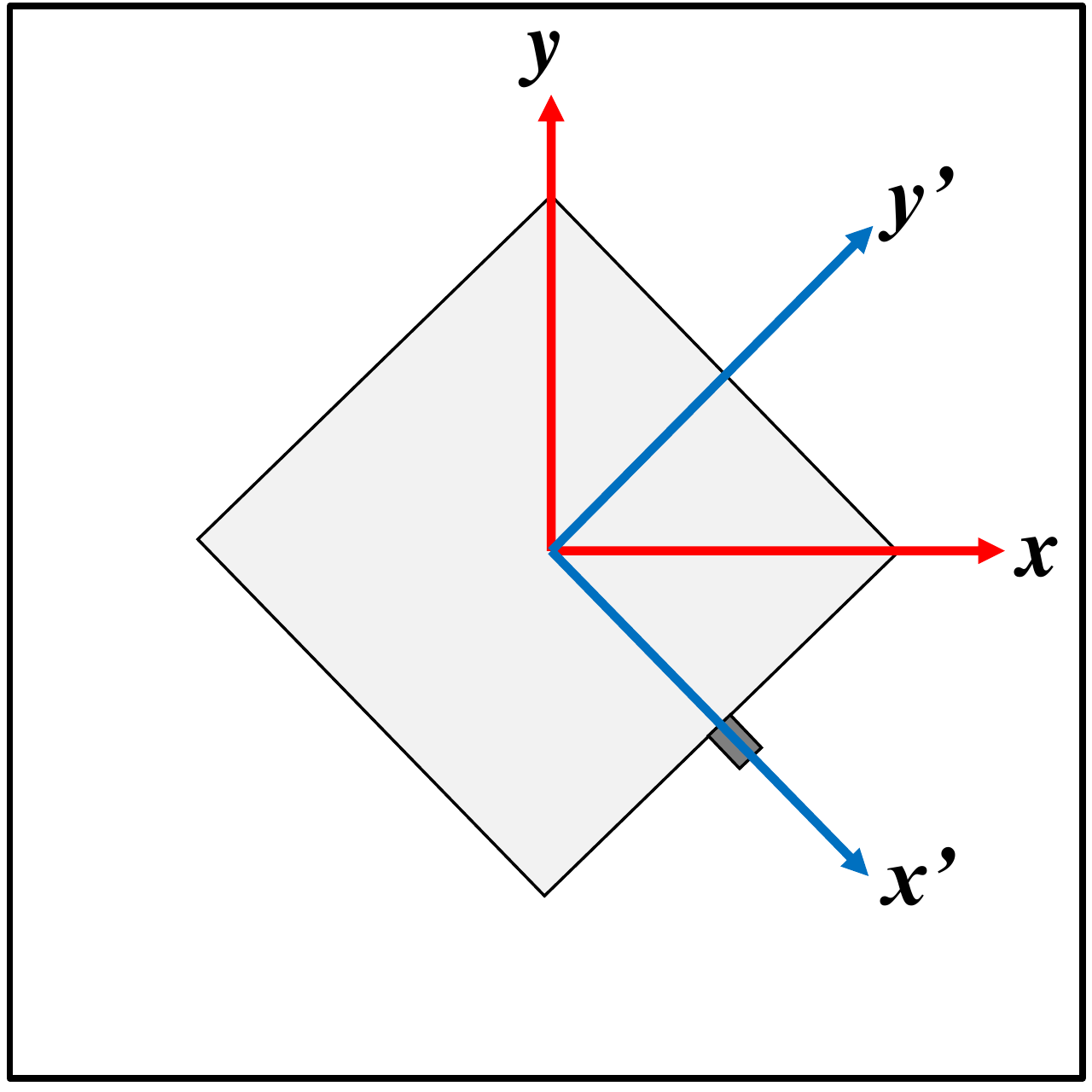}}
	\caption{Illustration of coordinate system definitions considering polarization.}
	\label{fig:polarization}
\end{figure}

Here, the principle of 1-bit polarization reconfiguration is reviewed in advance. Suppose the incident polarization is along $x$ axis. One switch is loaded along $x'$ axis, where $x'$-$y'$ coordinate system is rotated by 45$^\circ$ from $x$-$y$ coordinate system, as Figure. \ref{fig:polarization} shows. Suppose the scattering responses along $x'$ and $y'$ axes are isolated. If the additional phase along $y'$ axis is $\varphi_{y'}$ and the phase along $x'$ axis is $\varphi_{x'}$, under $x'$-$y'$ coordinate system, the element modulation matrix is simply written as
\begin{equation}
	\label{dualmodulation}
	\mathbb{M}'=\begin{bmatrix}
		e^{j\varphi_{x'}} & 0 \\ 0 & e^{j\varphi_{y'}}
	\end{bmatrix},
\end{equation}
and the incident wave is 
\begin{equation}
	\label{incwave}
	|p'\rangle=\frac{1}{\sqrt{2}}\begin{pmatrix}1 \\ 1\end{pmatrix}.
\end{equation}
Hence, the scattered wave is 
\begin{equation}
	\label{scatpie}
	\vec{E}'=\mathbb{M}' |p'\rangle
	=\frac{1}{\sqrt{2}}\begin{pmatrix}e^{j\varphi_{x'}} \\ e^{j\varphi_{y'}}\end{pmatrix}.
\end{equation}
Under $x$-$y$ coordinate system, the scattered wave is
\begin{equation}
	\label{esca}
	\vec{E}=\frac{1}{2}\begin{pmatrix}e^{j\varphi_{{x'}}}+e^{j\varphi_{{y'}}} \\ e^{j\varphi_{{x'}}}-e^{j\varphi_{{y'}}} \end{pmatrix}.
\end{equation}
Since the phase along $y'$ is fixed, suppose $\varphi_{y'}$ is fixed as 180$^\circ$; the phase along $x'$ is reconfigurable. Let us consider three special cases. (1) $\varphi_{x'}^{\rm{ON}} =180^\circ$, $\varphi_{x'}^{\rm{OFF}}=0^\circ$. According to (\ref{esca}), the scattered wave switches between LP($x$) and LP($y$). (2) $\varphi_{x'}^{\rm{ON}}=90^\circ$, $\varphi_{x'}^{\rm{OFF}}=0^\circ$. $\vec{E}^{\rm{ON}}$ is circular polarization, and $\vec{E}^{\rm{OFF}}$ is linear polarization. Hence, polarization modulation between LP and CP waves is obtained. (3) $\varphi_{x'}^{\rm{ON}}=90^\circ$, $\varphi_{x'}^{\rm{OFF}}=-90^\circ$. The scattered waves are two spins of CP. Therefore, dual-CP transition can also be generated and switched. The discussions are summarized in Table. \ref{tab:pol}. 

\begin{table}[!t]
	\centering
	\caption{A summary of 1-bit polarization manipulation types.}
	\scalebox{0.8}{
	\renewcommand\arraystretch{1.3}
	\begin{tabular}{|c|c|c|}
		\hline
		$\varphi_{y'}=180^\circ$     & Incident polairzation         & Scattered polarization                   \\ \hline
		$\varphi_{x'}^{\rm{ON}} =180^\circ$ & LP($x$) & LP($x$) - LP($y$) \\ \cline{2-3}
		$\varphi_{x'}^{\rm{OFF}}=0^\circ$ & CP & RHCP - LHCP \\ \hline
		$\varphi_{x'}^{\rm{ON}} =90^\circ$ & LP($x$) &	\multirow{2}{*}{LP-CP} \\ \cline{2-2}
		$\varphi_{x'}^{\rm{OFF}}=0^\circ$ & CP & \\ \hline
		$\varphi_{x'}^{\rm{ON}} =90^\circ$ & LP($x$) & RHCP - LHCP \\ \cline{2-3}
		$\varphi_{x'}^{\rm{OFF}}=-90^\circ$ & CP & LP($x$) - LP($y$) \\ \hline
	\end{tabular}}
	\label{tab:pol}
\end{table}

Some prototypes verify the transitions of LP($x$)-LP($y$), LP-CP and LHCP-RHCP in literatures \cite{summarypol,summary3pol,1bitpol1LPLP,1bitdirectionpolLPLP,1bitpol2LPCP,1bitpolCPCP,1bitphaseorpol}. LP-LP switching based on transmitarray is investigated in \cite{1bitpol1LPLP,1bitdirectionpolLPLP,1bitphaseorpol}. By tuning the ON/OFF states of PIN diodes, the reflection linear polarization is converted or remained (Figure. \ref{1bitsummary}(k)). Literature \cite{1bitpol2LPCP} studies the LP-CP switching type with LP incidence (Figure. \ref{1bitsummary}(l)). When the switch is ON, the reflection phase difference along $x'$ axis and $y'$ axis is 180$^\circ$, so the reflection polarization is converted. When the switch is OFF, the phase difference is 90$^\circ$, so CP is generated and reflected. Polarization-only R-MTS can also convert LP incident wave to RHCP/LHCP dynamically \cite{1bitpolCPCP}, if phase difference between $x'$ axis and $y'$ axis is 90$^\circ$ or -90$^\circ$ modulated by states of switches (Figure. \ref{1bitsummary}(m)). We note that other states of polarizations can also be switched as summarized in Table. \ref{tab:pol}.

\subsection{Direction-Only Element}
\label{directiononly}

In the formula (\ref{elementmodulation}), $F$ denotes the element radiation pattern, which determines the direction of the main beam scattered by a single element. It is worth mentioning that element radiation pattern is not exactly the array radiation pattern. The latter not only considers the element radiation pattern, but is also influenced by other dimensions of elements like phase and amplitude, as derived in (\ref{phase-only}) and (\ref{phase-only-invar}).

Reflection-transmission pattern reconfigurable AFSSs are comprehensively studied in recent years \cite{1bitdirectionpolLPLP,1bitdirection2,1bitdirection3,1bitdirection,1bitdirection4}. A dipole FSS with tunable PIN diodes at the center is reported in \cite{1bitdirection2}, which can switch the reflection/transmission states of EM wave (Figure. \ref{1bitsummary}(n)). If the PIN diode is at ON state, the dipole is resonant, so the EM wave is blocked by the FSS and then reflected. Otherwise, if the PIN diode is at OFF state, the dipole is not excited, so EM wave can pass through the FSS. 
Later, elements with complementary reflection and transmission responses in full polarization are reported in \cite{1bitdirection4}, as shown in Figure. \ref{1bitsummary}(o). The \#-shaped AFSS element can respond to full polarizations.

Note that $\vec{k}$ is a vector in full space, which means beyond propagating forward or backward. Propagating along different angles based on pattern reconfiguration is also direction-only types. Though reflection- or transmission-type direction reconfigurable elements are well studied, to the best of our knowledge, direction-only elements dedicated to pattern reconfiguration at one side (angle reconfiguration) are not reported in published literatures, and more related works could be expected in the future.

\subsection{Frequency-Only Element}
\label{frequencyonly}

The inherent operating mechanism of new frequency generation is nonlinear effect. Hence, EM elements should be nonlinear if the function of R-MTS is converting frequency. Frequency-only R-MTS has been under investigation since 1980s \cite{nonlinearsummary}. However, because of the complexity of elements and the lack of applications, researchers have paid little attention to this topic in last decades. Recently, the rapid development of digital controlling circuits like field programmable gate array (FPGA) makes it possible to use time dimension to generate new frequency based on R-MTS \cite{freqtime1,freqtime2,freqtime3,freqtime4,freqtime5}. Figure. \ref{1bitsummary}(p) shows a frequency-conversion element, which can mix frequencies of spatial EM wave and guided wave, and re-radiate the mixed wave into free space \cite{freqtime5}.

As shown in (\ref{Ri}), notation $R$ enables the reconfiguration of elment, as a function of time $t$. Suppose periodic controlling signal is applied to a phase-only element with period $T$. Then the additional phase of elements can be modulated periodically. According to formula (\ref{elementmodulation}), and considering the only variable is phase ($\Delta\varphi$), the main term of element modulation function is
\begin{equation}
	M(t) = e^{j\Delta\varphi(t)}.
\end{equation}
$M(t)$ is also a periodic funtion of time $t$, so applying the Fourier expansion, $M(t)$ is expressed as
\begin{equation}
	M(t) = \sum_{h=-\infty}^{\infty}a_h e^{j\frac{2\pi h}{T} t},
\end{equation}
where $\{a_h\}$ is the Fourier series, and
\begin{equation}
	\label{fourierseries}
	a_h = \frac{1}{T}\int_{-\frac{T}{2}}^{\frac{T}{2}} e^{j\Delta\varphi(t)} e^{-j\frac{2\pi h}{T} t} dt.
\end{equation}
Harmonic frequency components are generated, provided that the $\Delta\varphi(t)$ is not a constant number. According to formula (\ref{response}), when the incident frequency is $f_c$, the core term of scattered wave is 
\begin{equation}
	E(t) = \sum_{h=-\infty}^{\infty}a_h e^{j(2\pi f_c+\frac{2\pi h}{T}) t}.
	\label{timesca}
\end{equation}
As (\ref{timesca}) illustrates, new frequencies are added to the incident frequency, and the harmonic frequency components of spatial wave are reconfigurable if the controlling signal is reconfigurable or the switching period is changeable.

Essentially, EM elements act as a frequency mixer, mixing the frequency in free space and the switching frequency of FPGA on board. The reason why rapidly switching element can work as mixer results from the fact that the lumped switch is a nonlinear component. Temporal modulation of the ON/OFF states of switch generates square wave of phase, which is the inherent source of new frequency generation. We remark that the utmost frequency shift is determined by the switching speed of switches and FPGA, which is low so far compared to the frequency of spatial EM wave ($1/T\ll f_c$). So the significance of frequency shift realized by rapidly switching states on elements needs to be improved.

\section{2-Bit Element}
\label{secV}
This section reviews the advances of 2-bit elements with two independent switches. With more information bits manipulated by a single element, R-MTS goes to be multi-dimensional, and a variety of functions can be realized by R-MTS.

Note that the response notation $R$ in (\ref{Ri}) is determined by two parts: \textit{control} and \textit{excitation}. Following the information allocation strategy, there are two types of allocating 2 bits: both two bits go to \textit{control}, or one bit goes to \textit{control} and another bit goes to \textit{excitation}.

\begin{figure}[!t]
	\centering
	\subfigure[] { \label{2bit} 
		\includegraphics[width=0.45\columnwidth]{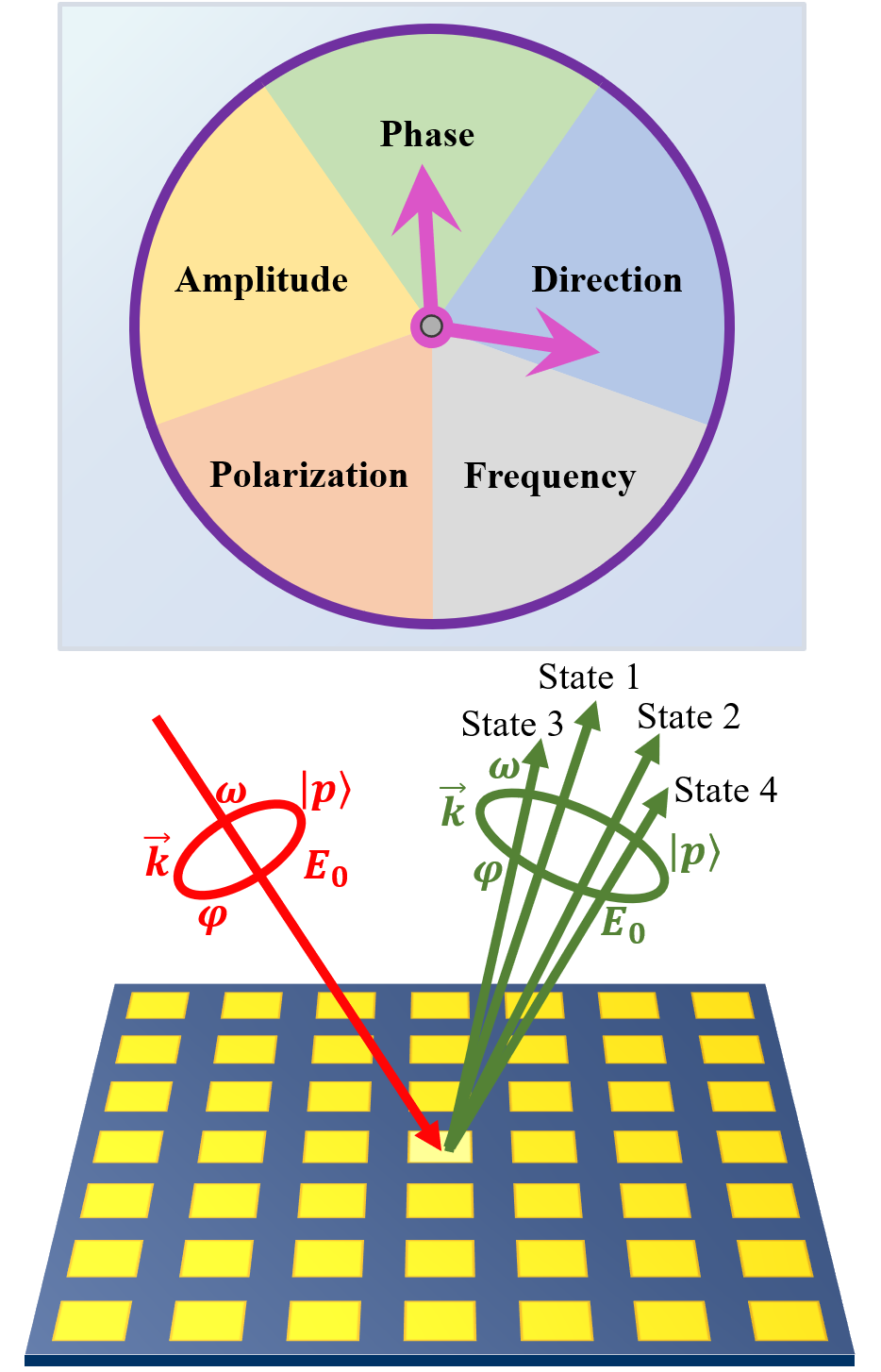} 
	} 
	\subfigure[] { \label{2bitmultiplex} 
		\includegraphics[width=0.45\columnwidth]{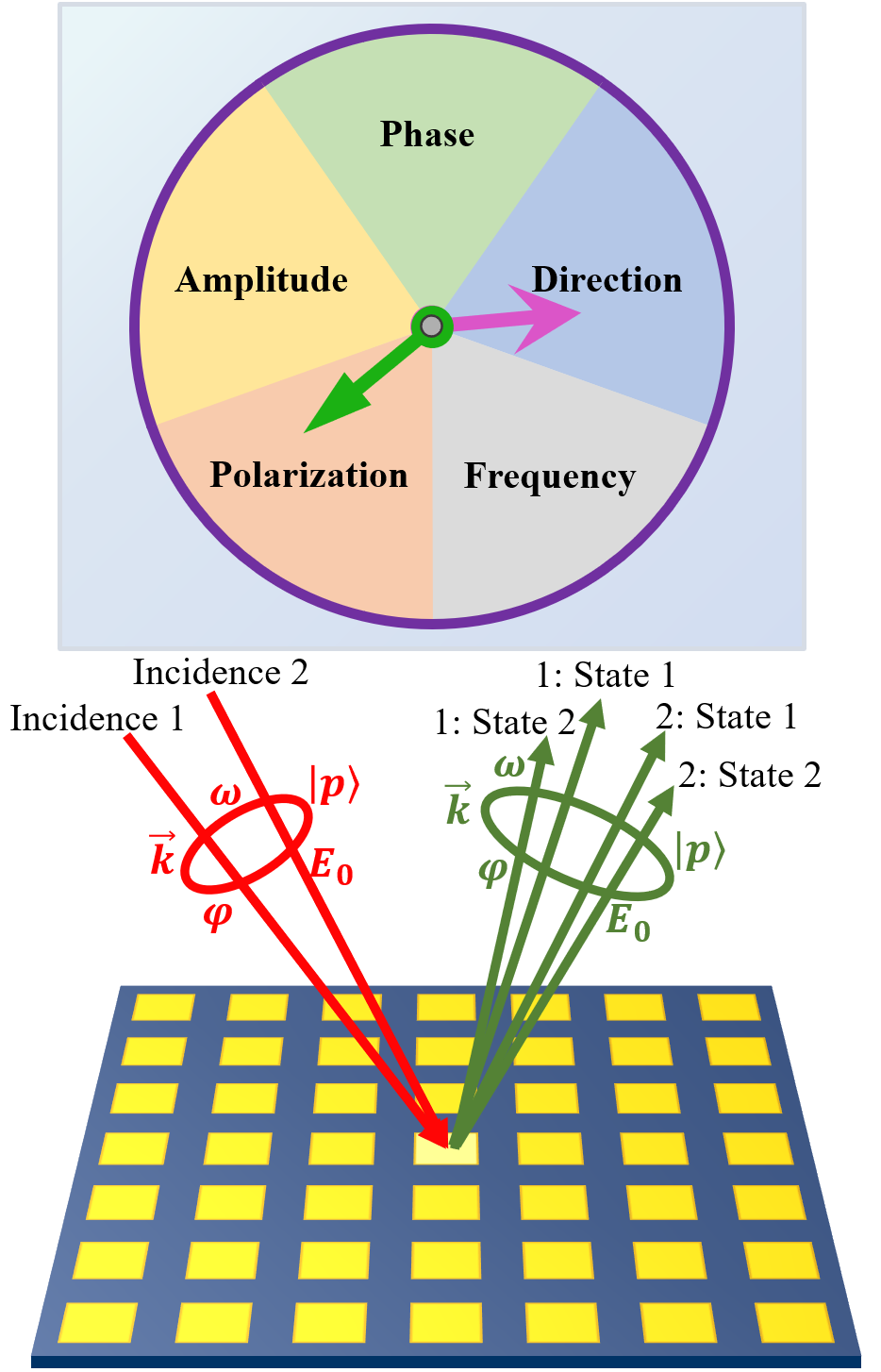} 
	} 
	\caption{Illustrations of two types of 2-bit allocation. (a) 2-bit-manipulating type, which can actively manipulate one or two dimensions under one incident wave. Two bits are allocated to five dimensions, so there are $C_5^2+5=15$ combinations of this type of bit allocation. (b) Multiplexed-manipulating type, which can manipulate one dimension (pink arrow) while responding to two incident waves differing in one dimension (green arrow). Theoretically, there are $5\times5=25$ ways of bit allocation of this type.}
	\label{2bitall}
\end{figure}

In the first type, each element can actively manipulate 2-bit information at one time, which is called 2-bit-manipulating type. Since there are five dimensions of EM wave, 2 bits go to these five dimensions respectively, as shown in Figure. \ref{2bit}. There are $C_5^2$ ways to allocate 2 bits to two different dimensions, and 5 ways to allocate all the 2 bits to one dimension. Totally, there are $C_5^2+5=15$ ways of bit allocation of the 2-bit-manipulating type elements.

In the second type, element can be designed to actively control 1 bit information on one dimension, while responding to excitations of two incident waves on one dimension independently using the other 1 bit. Two incident waves can be independently manipulated by multiplexing one element, so we call these elements multiplexed-manipulating type elements. Theoretically, there are $5\times5=25$ combinations of this type, as shown in Figure. \ref{2bitmultiplex}.

Here, some existing combinations are reviewed respectively, and the others still need to be further studied.

\subsection{2-Bit-Manipulating Type Element in One Dimension}

\begin{figure*}[!t]
	\centerline{\includegraphics[width=1.8\columnwidth]{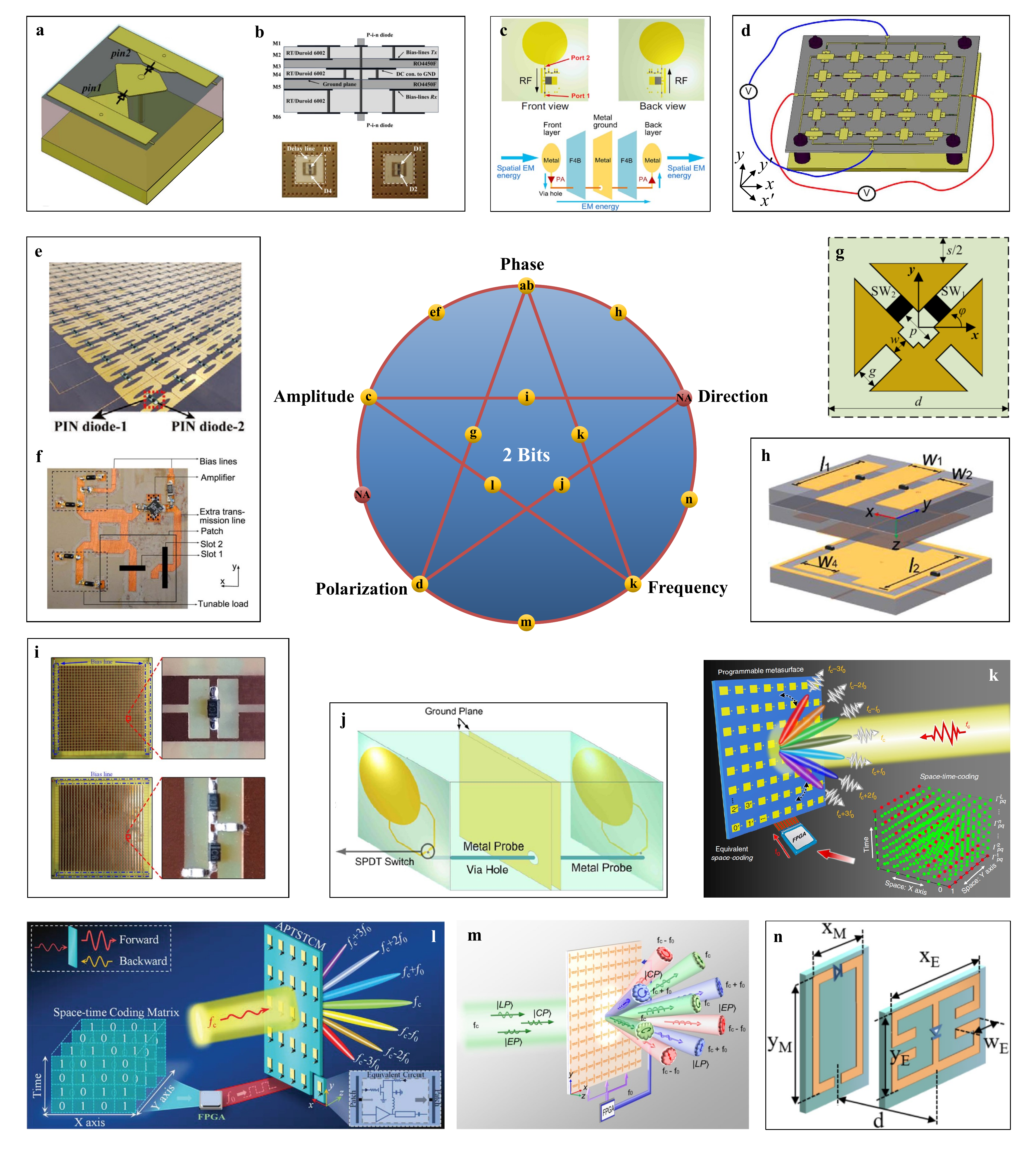}}
	\caption{2-bit reconfigurable elements actively manipulating one or two of five dimensions. 
		(a) 2-bit-phase reconfigurable reflectarray based on four switchable resonant states [71]. 
		(b) 2-bit-phase reconfigurable transmitarray based on receive-transmit structure with two sets of switches on each side [73]. 
		(c) A 2-bit-amplitude reconfigurable element realizing amplitude amplifying, maintaining and attenuating functions [77]. 
		(d) 2-bit-polarization R-MTS with the ability to switch among RHCP, LHCP and LP states [78]. 
		(e) A phase-amplitude R-MTS with one PIN diode modulating phase and the other controlling amplitude attenuating or not [80]. 
		(f) A phase-amplitude reconfigurable reflective element with varactor tuning the phase and transistor controlling the amplitude [81]. 
		(g) A phase-polarization reconfigurable element with two independent switches operating synchronously [83]. 
		(h) Phase-direction reconfigurable element with two PIN diodes modulating phase and two PIN diodes manipulating direction [87]. 
		(i) A direction-amplitude reconfigurable prototype [89]. 
		(j) Direction-polarization reconfigurable element using two SPDT switches [91]. 
		Polarization and direction are flexibly controlled. 
		(k) An illustration of frequency-phase reconfiguration [92]. 
		The beam directions of harmonic frequency waves are individually controlled and the harmonic frequencies are tuned. 
		(l) The frequency-amplitude R-MTS, where the amplification level is tunable at harmonic frequencies based on space-time modulation [97]. 
		(m) The frequency-polarization reconfiguration based on time modulation strategy, which can generate arbitrary polarization at harmonic frequencies [99]. 
		(n) A frequency-direction reconfigurable Huygens' element based on dynamic space-time modulation, where the ratio between forward and backward energy is tunable [100]. 
	}
	\label{2bitsummaryactive}
\end{figure*}

\subsubsection{2-Bit-Phase Reconfigurable Element}

2-bit-phase element with 2 bits both allocated to phase dimension is a well studied type \cite{2bitphase7,2bitphase5,2bitphase3,2bitphase6,2bitphase1,2bitphase4}. Four states of phase, 0$^\circ$, 90$^\circ$, 180$^\circ$ and 270$^\circ$, are available in 2-bit-phase elements, which can improve the phase resolution and reduce the quantization loss of RRA by 2.4 dB - 3 dB compared to 1-bit phase \cite{phasequan,phasequan2}.

Two switches are minimum for 2-bit phase modulation. Both reflective and transmissive 2-bit-phase R-MTSs are realized. A reflective asymmetric pentagon-shaped element based on resonance method is proposed in \cite{2bitphase7}, as shown in Figure. \ref{2bitsummaryactive}(a), which can generate 0$^\circ$, 90$^\circ$, 180$^\circ$ and 270$^\circ$ reflection phase states. For 2-bit-phase transmitarrays, more switches are needed in receive-transmit structure based transmitarrays \cite{2bitphase3}. 0$^\circ$/180$^\circ$ phase shift is realized in receiving patch, and 0$^\circ$/90$^\circ$ phase shift is achieved in transmitting structure, and then four states of phase are generated (Figure. \ref{2bitsummaryactive}(b)).

\subsubsection{2-Bit-Amplitude Reconfigurable Element}

A 2-bit-amplitude R-MTS with amplitude amplifying, maintaining and attenuating functions is implemented by controlling supply voltages of two amplifiers on each side of receive-transmit structure based transmitarray \cite{2bitamp}, as shown in Figure. \ref{2bitsummaryactive}(c). In fact, continuous amplitude reconfiguration can also be achieved by continuously tuning supply voltages.

\subsubsection{2-Bit-Polarization Reconfigurable Element}

2-bit-polarization reconfigurable elements are also reported in recent literatures. Three states of polarizations are actively generated and modulated using two independent switches \cite{2bitpol}. PIN diodes are soldered along $x, y$ axis respectively (Figure. \ref{2bitsummaryactive}(d)), with 90$^\circ$ ON/OFF phase difference along both axes. So 90$^\circ$, -90$^\circ$ and 0$^\circ$ phase differences between $x, y$ axes are available by independently tuning the two switches. If the incident LP is along $x'$ axis, RHCP, LHCP and LP reflection waves are obtained respectively. It is worth mentioning that the rest state of four states is the 90$^\circ$ phase of LP which is not used in this design. 

Furthermore,to realize four-state polarization reconfiguration of RHCP, LHCP, LP($x'$) and LP($y'$), we propose two approaches. It can be obtained with incident polarization along $x'$ if the additional phase is 0$^\circ$/180$^\circ$ along $x$ axis, and 0$^\circ$/90$^\circ$ along $y$ axis. Besides, another approach is that, if the reflection phase along $y$ axis is fixed such as 180$^\circ$, 2-bit-phase reconfigurable element with tunable phase along $x$ axis can also act as a 2-bit-polarization reconfigurable element, when incident polarization is along $x'$ axis.

\subsubsection{2-Bit-Frequency Reconfigurable Element}

Frequency reconfigurable elements are usually enabled by time modulation, which will be discussed in Section \ref{2bitfrequency}.

\subsection{2-Bit-Manipulating Type Element in Two Different Dimensions}

\subsubsection{Phase-Amplitude Reconfigurable Element}

If one bit is allocated to phase, and the other bit goes to amplitude, it is called phase-amplitude reconfigurable element.
Literature \cite{2bitphaseamp2} reports a dual-layer structure, with the top layer formed by graphene which manipulates the reflection amplitude, and the bottom layer composed of PIN diodes which tunes the reflection phase. In \cite{2bitphaseamp1}, single-layer elements integrated with two types of PIN diodes are proposed (Figure. \ref{2bitsummaryactive}(e)). Here, one PIN diode has little insertion loss at ON state and high isolation at OFF state, which is employed to tune reflection phase while maintaining the reflection amplitude. The other PIN diode also has little insertion loss at ON state, the, but at OFF state, it can absorb a large part of energy while impact little influence on the phase. Hence, a 2-bit phase-amplitude reconfigurable element is realized via these careful designs.

Amplifying reflectarray while modulating reflection phase is proposed in \cite{2bitphaseampref}, as shown in Figure. \ref{2bitsummaryactive}(f). A patch and I-shaped slot convert $y$-polarized spatial wave into guided wave. Varactors are loaded on transmission line, which can modulate the transmission phase. Lumped amplifier is employed to amplify the energy on transmission line. Then the $x$-polaried energy is re-radiated through slot and patch.

\subsubsection{Phase-Polarization Reconfigurable Element}

Phase-polarization reconfigurable elements have been designed and fabricated in early years. In 2010, literature \cite{2bitphasepol} introduced the idea of realizing phase modulation while controlling polarizations in reflectarrays. It can be achieved by dual-polarized reconfigurable element (we classify it as \textit{polarization-multiplexed phase-manipulating} type in Section \ref{pol-phase}) with incident polarization along $x'$ axis. The lossless modulation matrix of dual-polarized reconfigurable element in $x$-$y$ coordinate system is 
\begin{equation}
	\mathbb{M}=\begin{bmatrix}
		e^{j\varphi_{x}} & 0 \\ 0 & e^{j\varphi_{y}}
	\end{bmatrix}.
\end{equation}
Following the similar derivation in (\ref{incwave})-(\ref{esca}), the reflection wave in $x'$-$y'$ coordinate system is 
\begin{equation}
	\vec{E}'=\frac{1}{2}\begin{pmatrix}e^{j\varphi_{{x}}}+e^{j\varphi_{{y}}} \\ e^{j\varphi_{{x}}}-e^{j\varphi_{{y}}} \end{pmatrix},
\end{equation}
with 1-bit reconfigurable phase along $x$ and $y$ axes independently. Suppose the phases switch between 0$^\circ$/180$^\circ$ at OFF/ON states, four states of reflection wave are obtained as follows
\begin{equation}
	\begin{aligned}
		&\vec{E}'_{\rm{ON-ON}}=\begin{pmatrix} e^{j180^\circ} \\ 0 \end{pmatrix},\text{ } \vec{E}'_{\rm{OFF-OFF}}=\begin{pmatrix} e^{j0^\circ} \\ 0 \end{pmatrix},\\
		&\vec{E}'_{\rm{OFF-ON}}=\begin{pmatrix} 0 \\ e^{j0^\circ} \end{pmatrix},\text{ } \vec{E}'_{\rm{ON-OFF}}=\begin{pmatrix} 0 \\ e^{j180^\circ} \end{pmatrix}.
	\end{aligned}
\end{equation}
As it demonstrates, polarization switches dynamically between $x'$ and $y'$ axes, with independent modulation of the reflection phase.

A simplified reflective element with 2-bit phase-polarization reconfiguration is further designed and simulated in \cite{2bitphasepol2}, as shown in Figure. \ref{2bitsummaryactive}(g). In recent years, RTAs based on receive-transmit structure also demonstrate the capability of 2-bit phase-polarization modulation for LP($x$)-LP($y$) transition \cite{2bitphasepoltrans}. Furthermore, if the phase difference along $x'$ and $y'$ axes is designed as 90$^\circ$/-90$^\circ$, RHCP/LHCP can be generated and modulated, as demonstrated in \cite{2bitphasecpol4}.

\subsubsection{Phase-Direction Reconfigurable Element}

Phase-direction reconfigurable type emerges recently \cite{2bitphasedir1,2bitphasedir2, 2bitphasedir3}. For example, as shown in Figure. \ref{2bitsummaryactive}(h), the element is composed of two layers, with the top layer controlling phase and the bottom layer controlling direction. Each element is controlled independently with 1 bit for phase reconfiguration and 1 bit for direction reconfiguration.

\subsubsection{Direction-Amplitude Reconfigurable Element}

Direction-amplitude manipulation type is another combination of 2-bit elements for reflection, transmission or absorption switching applications \cite{2bitampdirection2,2bitampdirection}. In literature \cite{2bitampdirection2}, two layers of the substrate are both soldered with PIN diodes (Figure. \ref{2bitsummaryactive}(i)). PIN diodes on the top layer can determine whether the EM wave passes through the first layer or is absorbed. PIN diodes on the bottom layer can control the reflection or transmission states. Note that there are only three states of the 2-bit element, because if the EM energy is absorbed on the top layer, it is meaningless to consider the propagation directions of EM wave.

\subsubsection{Direction-Polarization Reconfigurable Element}

Direction-polarization reconfigurable element is also reported in a recent literature. A reflecting-transmitting R-MTS controlled by incident polarization is proposed in \cite{2bitpoldirection}. Two identical structures are designed on the top and bottom layers with a single-pole double-throw (SPDT) switch on each layer (Figure. \ref{2bitsummaryactive}(j)). The top SPDT switch can control which polarization state could be transmitted to the bottom layer, and the bottom SPDT switch can determine which polarization state of transmitting wave could be excited. Hence, if the incident polarization is along $x$ axis, three states are obtained: $x$-polarized reflection wave, $x$-polarized transmission wave and $y$-polarized transmission wave, with polarization and direction manipulated independently.

\subsection{2-Bit-Manipulating Type Element Related with Frequency Reconfiguration}
\label{2bitfrequency}

Frequency reconfiguration has been discussed in Section \ref{frequencyonly}. Here, exploiting properties of controlling signals can modulate frequency as well as other dimensions of EM wave. According to (\ref{fourierseries}), $a_h$ is a complex number representing the amplitude and phase responses of the $h$th harmonic frequency, determined by controlling signal waveform. Each $a_h$ can be designed with great degrees of freedom. Therefore, various wave dimensions at harmonic frequencies can be independently modulated by applying different switchable signal waveforms. 

\subsubsection{2-Bit-Frequency and Frequency-Phase Reconfigurable Element}

Literature \cite{2bitphasefreqFPGA,2bitphaseampfreqFPGA,2bitphaseampfreq3FPGA,2bitphasefreqdireFPGA,nbitampphasetime} demonstrate that phases at different harmonic frequencies can be independently modulated using 1-bit or 2-bit phase reconfigurable elements by carefully designing the phase waveform of elements. Hence, scanning beams at different frequencies can be generated simultaneously (Figure. \ref{2bitsummaryactive}(k)). Besides, the signal period $T$ can modulate the harmonic frequency, so the frequency is also reconfigurable.

\subsubsection{Frequency-Ampitude Reconfigurable Element}

Apart from frequency-phase reconfiguration, frequency-amplitude reconfiguration type is also reported recently \cite{2bitampfreq}. As shown in Figure. \ref{2bitsummaryactive}(l), an amplifier and a FET are loaded on the passive structure, and by dynamically controlling the FET, the R-MTS can generate tunable amplified EM wave at different harmonic frequencies.

\subsubsection{Frequency-Polarization Reconfigurable Element}

Recent years have also witnessed the development of arbitrary polarization reconfigration based on time-modulation strategy \cite{nbitfreqpol,2bitpolfreq}. For example, orthogonal dipoles with two sets of PIN diodes are modulated dynamically to change the transmitted amplitude and phase of orthogonal LP waves at harmonic frequenies (Figure. \ref{2bitsummaryactive}(m)), and thus the arbitrary transmitted polarization is generated by composing the two modulated LP waves. 

\subsubsection{Frequency-Direction Reconfigurable Element}

Frequency-direction reconfiguration type appears in \cite{2bitdirfreq}, as shown in Figure. \ref{2bitsummaryactive}(n). A time-varying Huygens' metasurface is designed to change the energy ratio between reflection and transmission at harmonic frequencies. By applying space-time modulation strategy, the beam scanning angles can also be manipulated.

\subsubsection{Discussion}

According to information allocation strategy, a single reconfigurable element with at least N-bit physical devices can modulate N-bit information. Here, by rapidly switching element states, some extra tunable bits are provided by time-domain modulation. Hence, this type of R-MTS is also called \textit{time-modulated metasurface}. Besides, when extending elements to form a periodic array, with different states at different element locations, the amount of information is also multiplexed, which is provided by space-domain modulation. This type is the \textit{space-modulated R-MTS}, which is the array level reconfiguration as discussed previously. Moreover, by combining space modulation and time modulation, space-time-modulated R-MTSs can exploit additional dimensions in both space and time domain, so they can implement novel functions, which is a promising topic for future studies.

\subsection{Multiplexed-Manipulating Type Element}
\label{MultiplexedType}

\begin{figure*}[!t]
	\centerline{\includegraphics[width=1.5\columnwidth]{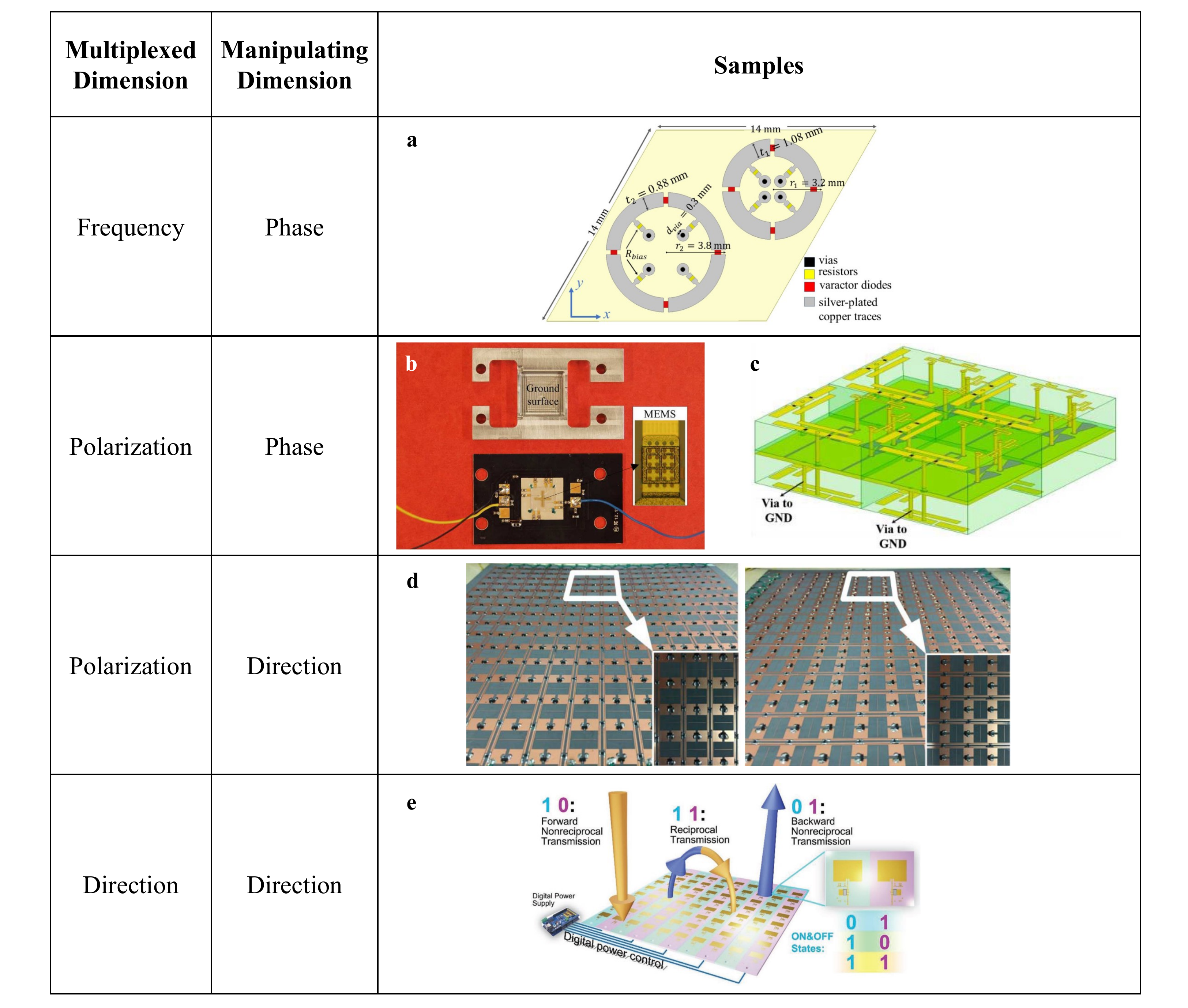}}
	\caption{2-bit reconfigurable elements with one multiplexed dimension and one manipulating dimension. 
		(a) Frequency-multiplexed phase-manipulating element with two reconfigurable elements in one supercell operating at the two frequency bands [102]. 
		(b) Polarization-multiplexed phase-manipulating reflective element [103]. 
		(c) Polarization-multiplexed phase-manipulating transmitarray [107]. 
		(d) A polarization-multiplexed direction-manipulating prototype [113]. 
		(e) A direction-multiplexed direction-reconfigurable design [116]. 
		The supercell can determine whether EM wave from two sides can pass through the MTS or not. 
	}
	\label{2bitsummarymultiplex}
\end{figure*}

Apart from active manipulation of two dimensions, actively manipulating one dimension while independently responding to two different incident waves is also a large category of 2-bit elements. There are 25 combinations of the multiplexed-manipulating type theoretically, but some are not practical. Here, we review the existing combinations, and propose some ideas of the unrealized combinations.

\subsubsection{Frequency-Multiplexed Phase-Manipulating Element}

Frequency-multiplexed phase control is a natural idea for full duplex communication using a single R-MTS. To independently modulate phase in response to two frequency bands with the same polarization, one can design a supercell consisted of two independent elements working at two bands \cite{2bitphasemulfreq1,2bitphasemulfreq}, as shown in Figure. \ref{2bitsummarymultiplex}(a) for an example.

\subsubsection{Polarization-Multiplexed Phase-Manipulating Element}
\label{pol-phase}
Polarization-multiplexed phase-manipulation is a well studied type \cite{2bitphasemulpol1,2bitphase17mulpol,2bitphasemulpol,2bitphasemulpol1freq,2bitphasemulpol2trans}. Elements based on isolation between two orthogonal directions can independently manipulate two polarizations. So a single element can work for two incident waves distinguished by orthogonal polarizations simultaneously. As shown in Figure. \ref{2bitsummarymultiplex}(b), a cross-shaped element with two sets of independent MEMSs can modulate reflection phase of dual-LP waves \cite{2bitphasemulpol1}. 

Additionally, dual-LP-multiplexed transmitarray is demonstrated in \cite{2bitphasemulpol2trans}. Two 1-bit phase reconfigurable dipoles are placed orthogonally to form an element, thus the transmitting polarization is multiplexed (Figure. \ref{2bitsummarymultiplex}(c)). 

The above mentioned literatures are all based on LP incidence. We notice that fixed MTSs for independent dual-CP multiplexing are sufficiently studied in \cite{unreconfigure1phasemulCP,unreconfigurephase3mulCP,unreconfigurephasemulCP1,unreconfigurephasemulCP}. However, as far as we know, independent phase reconfiguration for dual-CP incident wave is not reported in literatures, which might be a future research topic.

\subsubsection{Polarization-Multiplexed Direction-Manipulating Element}

Polarization-multiplexed direction-manipulation evolves from 1-bit dierction reconfiguration as demonstrated in Section \ref{directiononly}. Literature \cite{2bitdirectionmulpol2,2bitdirectionmulpol3} load orthogonal switches on two layers to manipulate propagating directions of two incident polarizations independently, shown in Figure. \ref{2bitsummarymultiplex}(d) as an example. Furthermore, polarization conversion layer is added on the top of these layers to implement polarization conversion functions \cite{2bitdirectionmulpol4}. Besides, fixed-phase layers are added on the top and bottom, and different holographic images are obtained from different incident polarizations with controllable directions \cite{2bitdirection1mulpol}.

\subsubsection{Direction-Multiplexed Direction-Manipulating Element}

A 2-bit R-MTS that modulates direction of wave while responding to two directions of incident wave appears in \cite{2bitdirectionmuldirection}. It exploits the inherent nonreciprocity of amplifier to realize the independent direction modulation of both sides. A supercell combined by two opposite amplifying elements can respond to forward and backward wave independently (Figure. \ref{2bitsummarymultiplex}(e)). If the forward element operates at ON state, the forward EM energy is amplified and transmitted; otherwise, the forward energy is blocked and reflected backward. The backward element can modulate the direction of the backward-incident EM wave similarly. Consequently, the incident waves from two directions are modulated independently.

\subsubsection{Unrealized Multiplexed-Manipulating Types to be Explored}

As demonstrated above, multi-functional R-MTSs of multiplexed-manipulating types are emerging nowadays. Here, some ideas of other multiplexed-manipulating type reconfigurable elements will be discussed in the next few paragraphs.

Direction-multiplexed phase-modulation is also a promising topic. Here, we discuss two types of direction multiplexing. The first type is called \textit{Janus metasurface} with reference to \cite{Janus,Janus1}. With Janus metasurface responding to bidirectional incident waves respectively, two unrelated transmitting holographic images are obtained from both sides of the MTS \cite{Janus1}. 

Besides, the term \textit{direction} is not only for the opposite directions, but also for the incident angles. Considering this, another type is angle-multiplexed phase modulation, where independent designed beam patterns are generated with different incident angles on the MTS. Angle-sensitive elements are designed in \cite{unreconfigurephasemulangle4,unreconfigurephasemulangle3,unreconfigurephasemulangle1,unreconfigurephasemulangle,unreconfigurephasemulangle5}, which can respond to different incident angles independently.

Direction-multiplexed MTSs are a new research trend nowadays, and all the above mentioned works focus on fixed MTSs. Reconfigurable direction-multiplexed phase-tuning could have great potential in the future.

Polarization-multiplexed reconfiguration is also an attractive topic. Besides phase reconfiguration and direction reconfiguration as reviewed above, polarization-multiplexed amplitude or frequency reconfiguration are unexplored topics. Related works could be carried out in the future.

\section{Emerging Topics and Future Trends}
\label{secVI}

\subsection{N-Bit Element}
\label{nbitelement}

\begin{figure}[!t]
	\centerline{\includegraphics[width=0.5\columnwidth]{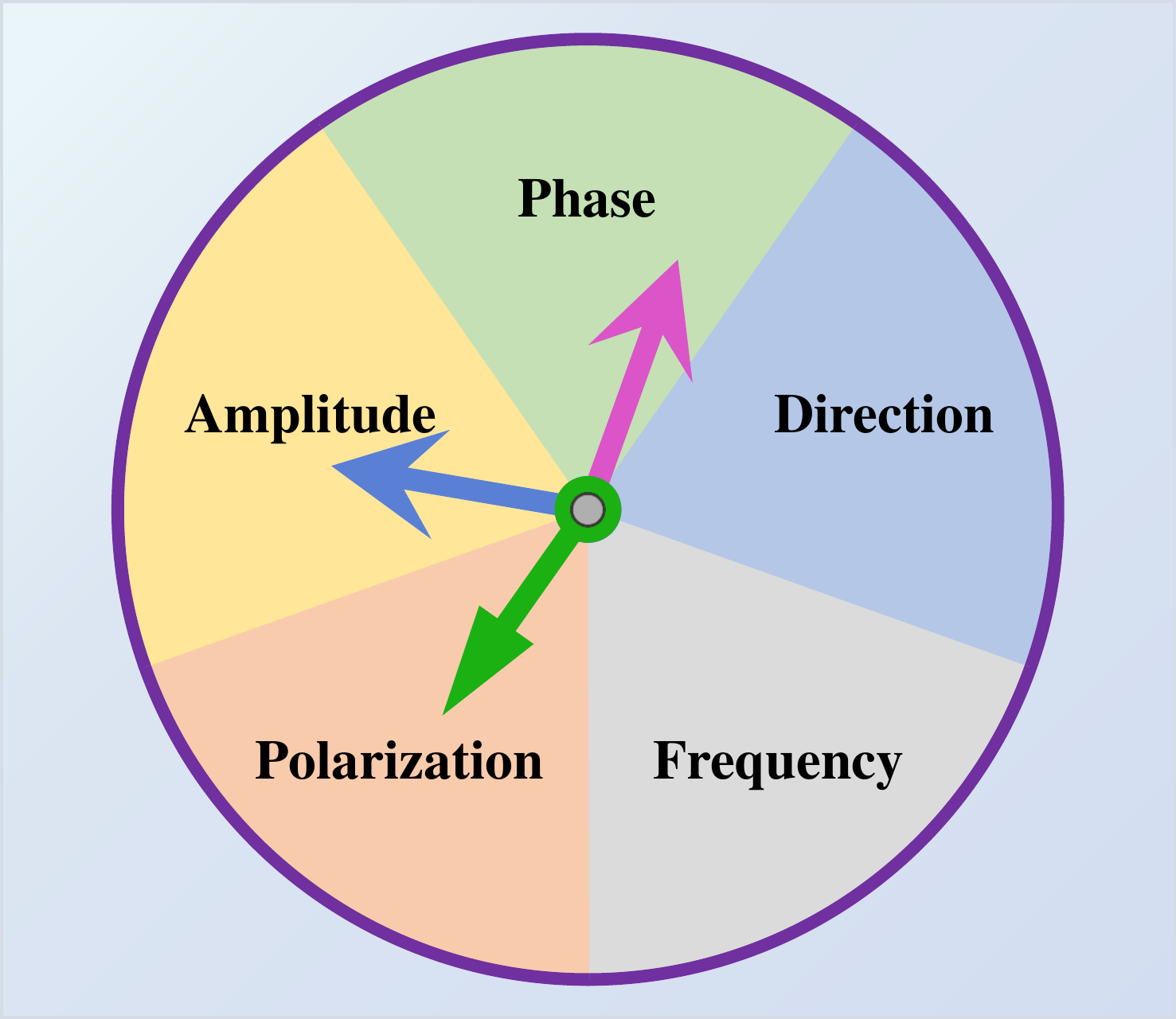}}
	\caption{An illustration of N-bit allocation. $N$ bits can be allocated to five dimensions independently with various combinations of multiplexed bits and manipulating bits. More dimensions are linked by a single element with higher resolution, so the functions of R-MTS can be greatly enriched compared with 1-bit or 2-bit R-MTS.}
	\label{nbit}
\end{figure}

This paper mainly reviews the 1-bit and 2-bit reconfigurable elements. Besides, elements with more than 2-bit reconfiguration are also one of the research highlights \cite{3bitphase2pol1,nbitphase}. Evolving from 1-bit and 2-bit allocation, an illustration of N-bit allocation is shown in Figure. \ref{nbit}.

Allocating $N$ bits to one dimension can improve the resolution on the dimension, thus improving the performance.
Furthermore, full-dimensional element with the ability to modulate five dimensions of EM wave can increase the functions of R-MTS, which is also a promising topic. We have known that the quantization loss of 2-bit phase reconfiguration is 0.6 dB - 0.9 dB \cite{phasequan,phasequan2}, which is acceptable in most applications. Conditions are similar for other dimensions. Therefore, allocating $N$ bits to diverse dimensions is generally more rewarding than just increasing the resolution of only one dimension. Besides, multiplexing a single element in multi-dimensions, while manipulating the multi-dimensional EM wave calles for N-bit multiplexed-manipulating type R-MTSs. Nonetheless, elements with more than 2-bit reconfigurable dimensions are insufficiently studied. 

To realize N-bit element, integrating more lumped switches on a single element is a natural approach. However, suffered from the bulky size of switches and the complexity of bias lines, the number of switches on a single element is limited. Applying continuously tunable switches like varactors can enable multi-bit reconfiguration too. However, this approach requires precise voltage control, which is not robust and increases the complexity of controlling circuit. 
As discussed in Section \ref{2bitfrequency}, time-modulated MTS has potential for N-bit modulation and multi-dimensional elements. Nevertheless, the limited radiation efficiency, complex controlling voltages, the narrow bandwidth and low modulation speed are the bottlenecks of this approach. Accordingly, multi-dimensional N-bit elements are still a challenging topic for future investigations.

\subsection{Terahertz and Optical R-MTS}
\label{highfreq}

Since high-performance lumped switches such as PIN diodes and varactors are available in microwave band, and the mainstream fabrication process like printed circuit board (PCB) technology is mature and low-cost, microwave band R-MTSs have experienced great breakthrough in recent decades. The demands for THz and optical R-MTSs are also emerging rapidly. However, in THz and optical band, commercial lumped switches are too large to be soldered and ON/OFF ratio is reduced. Scientists are seeking for new switches and new architectures with higher frequency and higher performance. With the great revolution of micro- and nano-fabrication technology in recent years, R-MTSs in THz and optical bands are experiencing rapid development nowadays. 

\begin{figure}[!t]
	\centering
	\subfigure[] { \label{fig:GaAs} 
		\includegraphics[width=0.45\columnwidth]{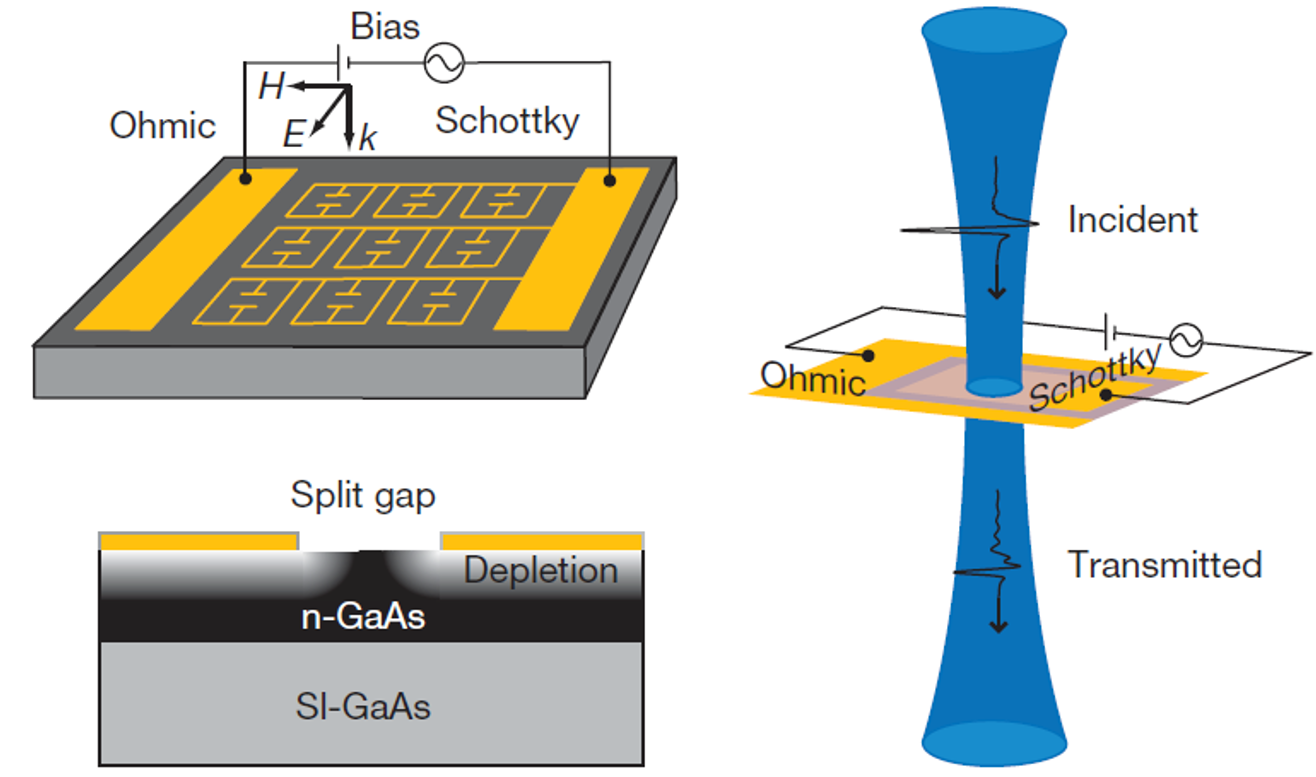} 
	} 
	\subfigure[] { \label{fig:1bitamp2THz} 
		\includegraphics[width=0.45\columnwidth]{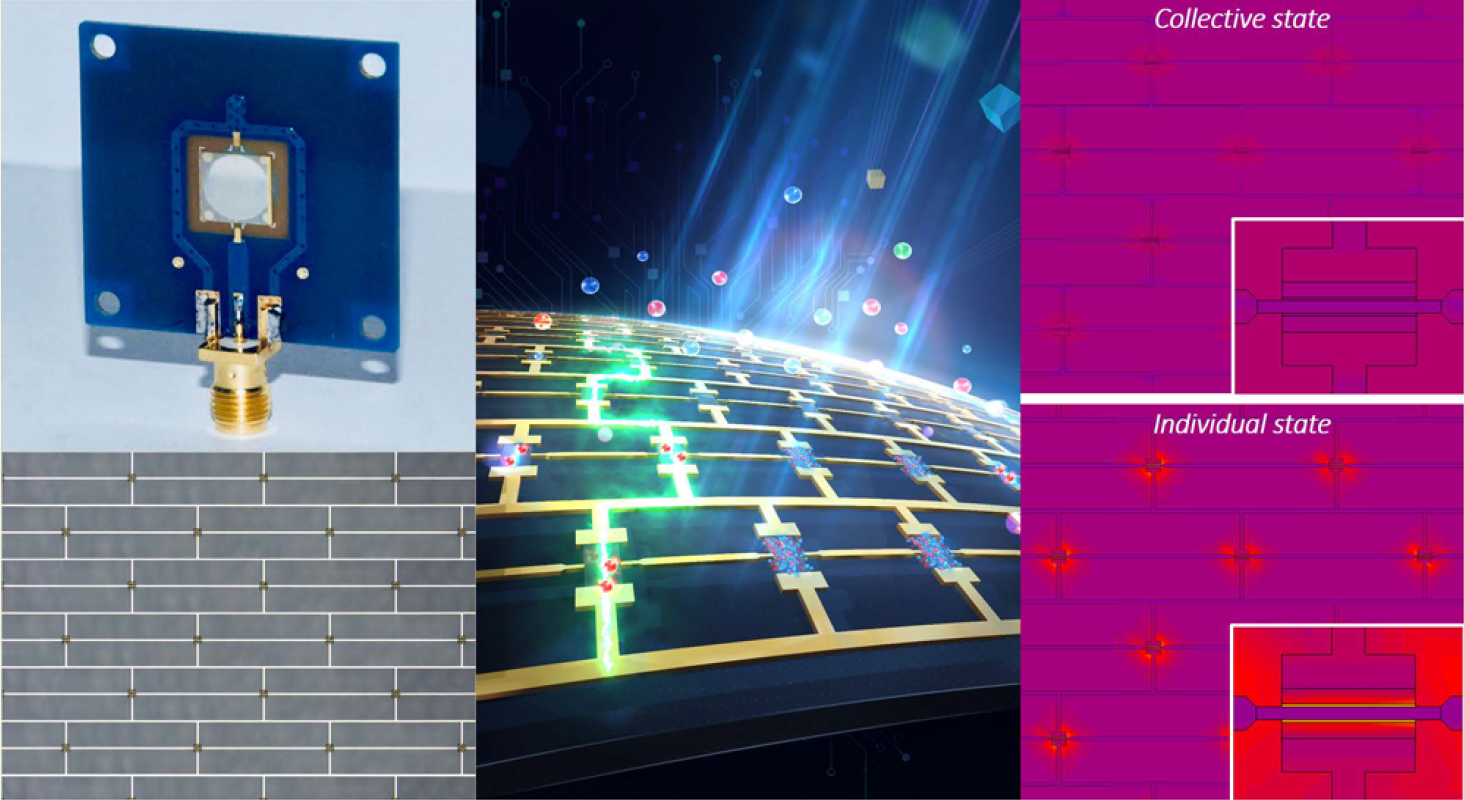} 
	} 
	\subfigure[] { \label{fig:1bitphase16THz} 
		\includegraphics[width=0.35\columnwidth]{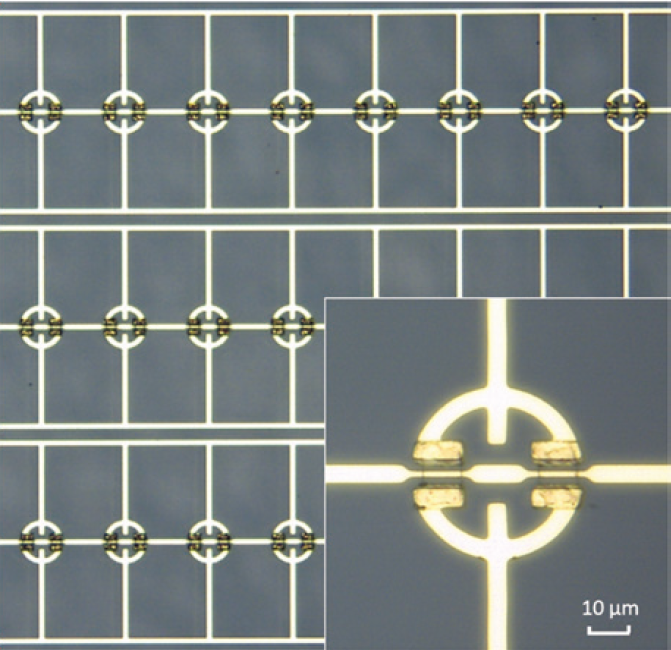} 
	} 
	\subfigure[] { \label{fig:nbitphase} 
		\includegraphics[width=0.38\columnwidth]{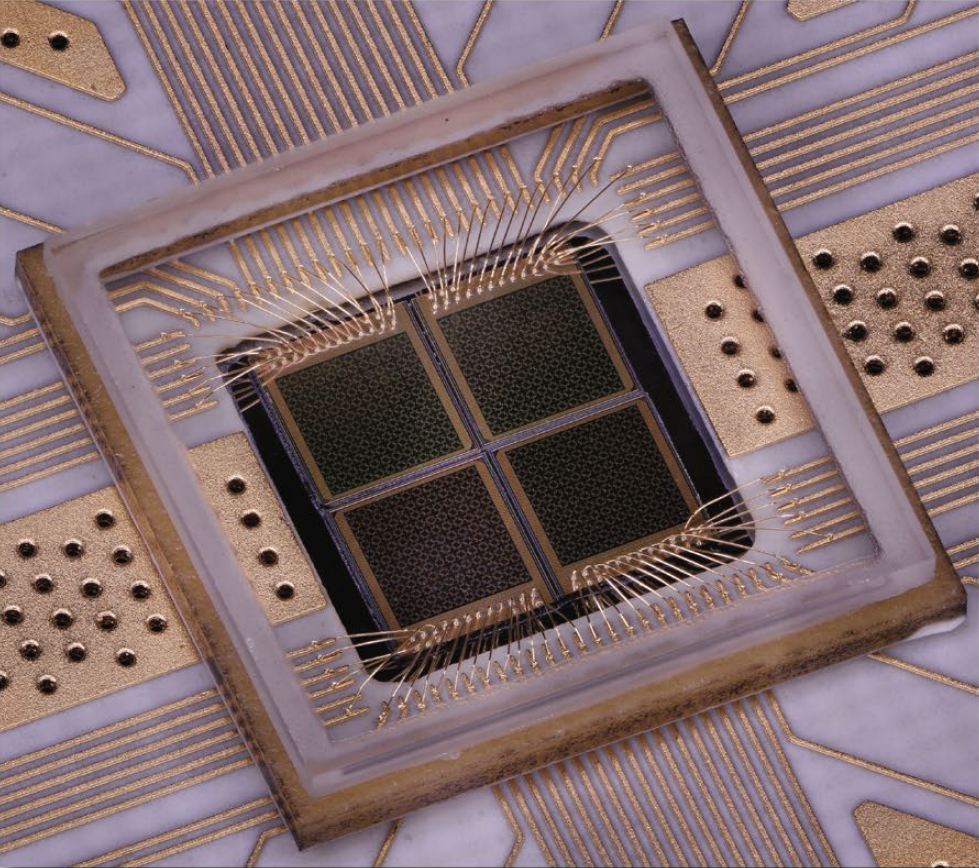} 
	} 
	\subfigure[] { \label{fig:1bitphaseopt1} 
		\includegraphics[width=0.45\columnwidth]{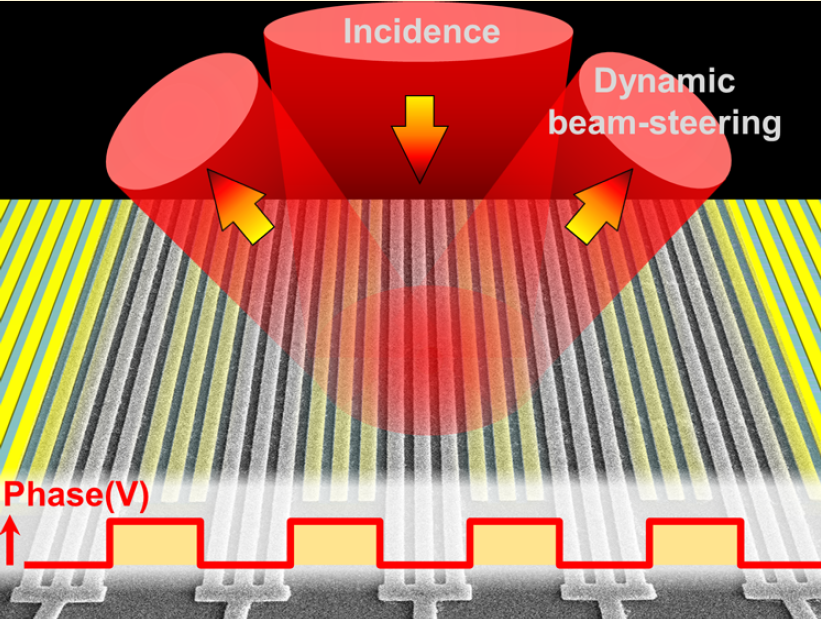} 
	} 
	\subfigure[] { \label{fig:1bitampopt2} 
		\includegraphics[width=0.45\columnwidth]{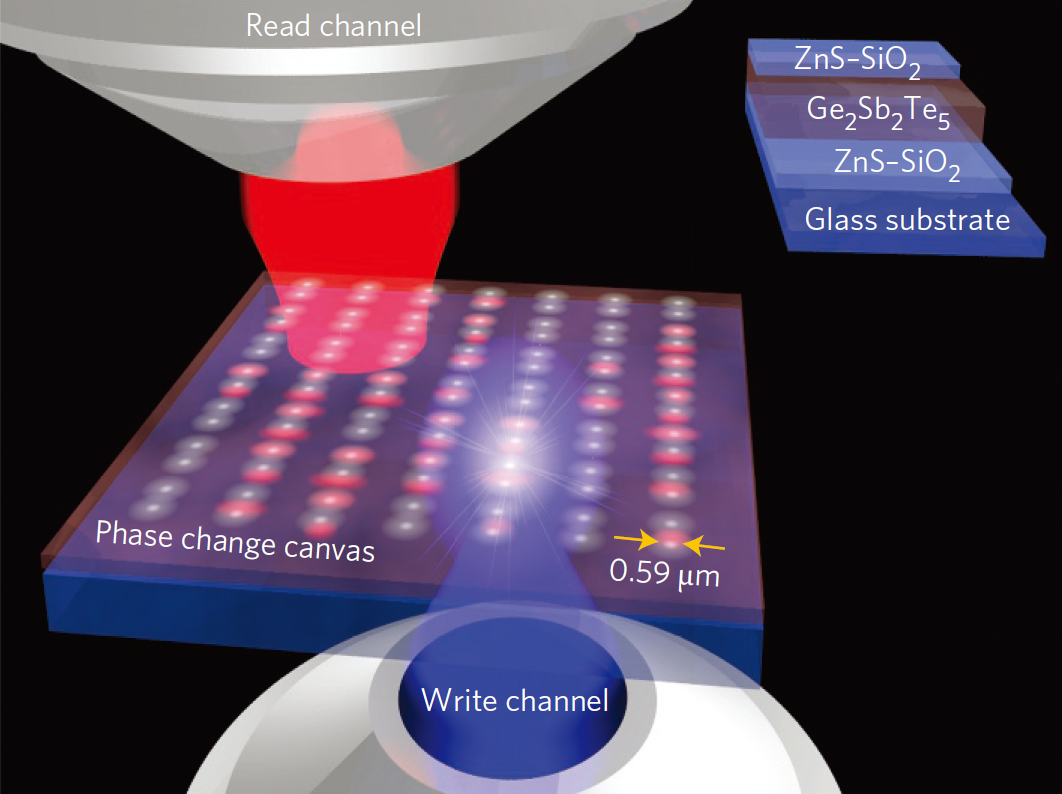} 
	} 
	\subfigure[] { \label{fig:1bitphaseopt5} 
		\includegraphics[width=0.7\columnwidth]{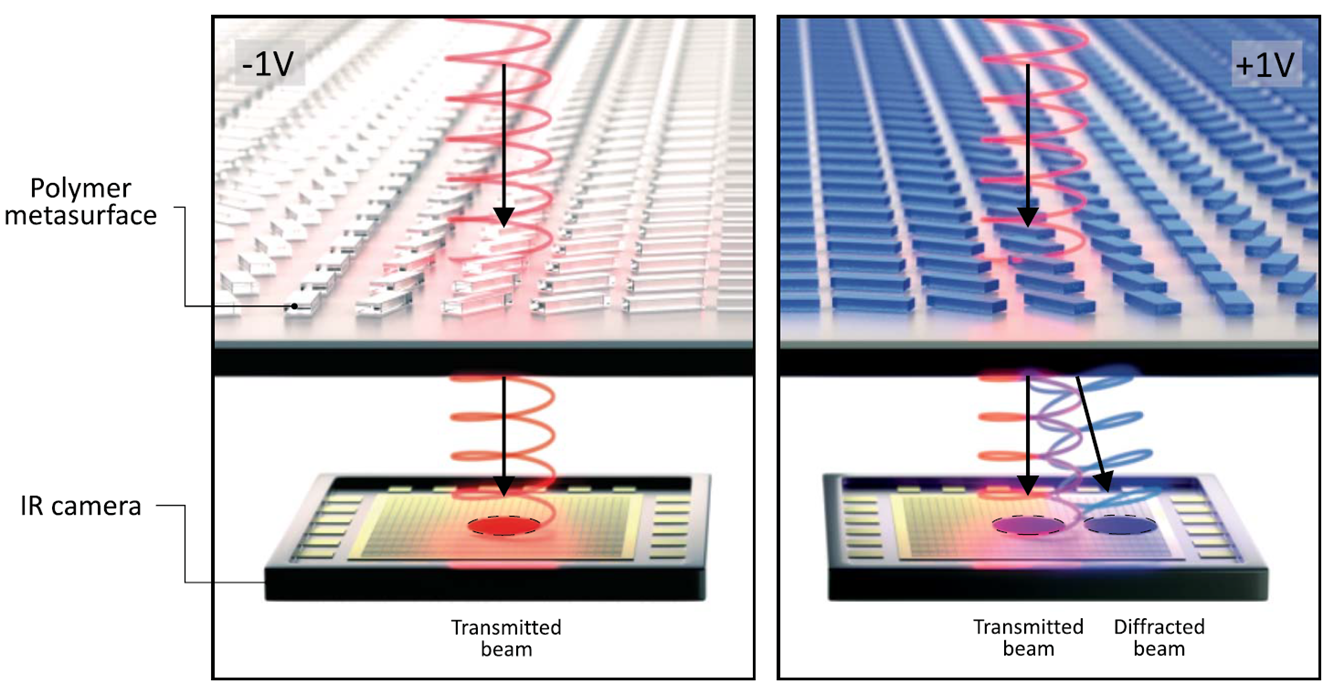} 
	} 
	\caption{(a)-(d) Terahertz and (e)-(g) optical R-MTSs. (a) Schottky diode based THz amplitude reconfigurable array \cite{GaAs}. (b) HEMT based THz amplitude reconfigurable transmitarray \cite{1bitamp2THz}. (c) HEMT based THz phase reconfigurable prototype \cite{1bitphase16THz}. (d) CMOS based THz phase reconfigurable prototype \cite{nbitphase}. (e) ITO based optical phase reconfigurable prototype \cite{1bitphaseopt1}. (f) GST based optical amplitude reconfigurable design \cite{1bitampopt2}. (g) PEDOT:PSS based optical phase reconfigurable design \cite{1bitphaseopt5}.}
	\label{THzoptical}
\end{figure}

In THz band, numerous switches are introduced, like Schottky diode \cite{GaAs,GaAs1} (Figure. \ref{THzoptical}(a)), high electron mobility transistor (HEMT) \cite{1bitphase19THz,1bitphase16THz,1bitamp1THz,1bitamp2THz,HEMT} (Figure. \ref{THzoptical}(b)-(c)), graphene \cite{graphene1,graphene}, complementary metal oxide semiconductor (CMOS) \cite{nbitphase,CMOS,CMOS1} (Figure. \ref{THzoptical}(d)), vanadium dioxide (VO$_2$) \cite{vo2,vo21,vo22}, to name a few. For instance, HEMT switch has high ON/OFF ratio and low controlling complexity, which is promised to provide 1-bit phase control in THz band \cite{1bitphase19THz,1bitphase16THz}. Besides, spatial THz amplitude modulators based on HEMT switches are reported in \cite{1bitamp1THz,1bitamp2THz}, with 93\% amplitude modulation depth and 1 GHz modulation rate.

In optical frequency band, no lumped switch is available neither, so element acts as both nanoantenna and switch \cite{1bitsummaryopt}. These switches are made of indium tin oxide (ITO) \cite{1bitphaseopt1} (Figure. \ref{THzoptical}(e)), graphene \cite{graphene3}, chalcogenide compound germanium-antimony-tellurium (GST) \cite{1bitampopt2,1bitampopt1,1bitampopt3,1bitphaseopt4,1bitphaseopt2,1bitphaseopt3} (Figure. \ref{THzoptical}(f)), liquid crystall \cite{liquidcry}, polymer poly(3,4-ethylenedioxythiophene):polystyrene sulfonate (PEDOT:PSS) \cite{1bitphaseopt5} (Figure. \ref{THzoptical}(g)) and so on. For example, phase-change materials like GST can provide obvious switching performance in optical band. Femtosecond pulses induce the amorphous-crystalline transition of GST \cite{1bitampopt2}, so the element resonance states are changed, and then direction, phase or amplitude can be modulated.

However, THz and optical R-MTSs still have many challenges to be handled, like improving efficiency, increasing switching speed, eliminating grating lobes and addressing independently. Additionally, multi-dimensional and multi-functional R-MTS in THz and optical bands are also the future trends.

\subsection{Surface-Wave R-MTS}

Above mentioned most MTSs manipulate spatial wave, where EM wave propagates in free space. Moreover, MTS also shows the ability of manipulating surface wave, where EM wave propagates along the tangential direction of MTS.

In optical frequency band, 2-D photonic crystals have been proposed to manipulate surface light since 1990s \cite{phtcry,phtcry1,phtcry2,phtcry3}, which is still a hot topic nowadays. To guide surface wave with more flexibility, photonic topological insulators \cite{topoins,topoins3,topoins1} are made artificially, which can control the light propagation directions on surface.

Periodic structures can also manipulate surface wave in microwave band. High impedance surfaces (HIS) form electromagnetic band gap (EBG) \cite{HIS,EBG1,EBG2,EBG3} to forbid wave propagation along MTS.

MTSs inducing the transition between surface wave and spatial wave are also studied. Phase-gradient metasurface can convert spatial EM wave to surface wave \cite{PGS1}. Besides, artificial impedance surface (AIS) can transform surface wave into directional spatial beams \cite{AIS}.

\begin{figure}[!t]
	\centerline{\includegraphics[width=0.8\columnwidth]{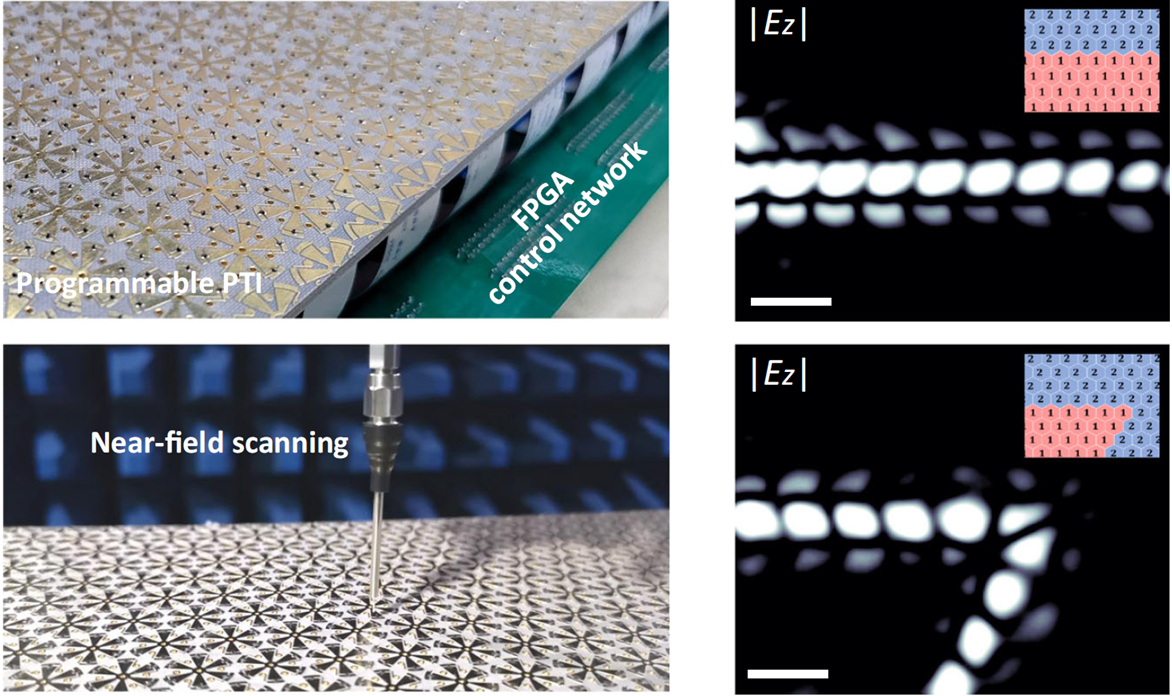}}
	\caption{Reprogrammable topological insulator in microwave band \cite{reconfigEBG}. The propagating directions on surface are programmed by independently addressing the configurations of elements.}
	\label{fig:reconfigEBG}
\end{figure}

Despite great breakthroughs in surface wave manipulation, R-MTSs for surface wave modulation are less investigated than the fixed surface-wave MTSs. A few R-MTSs are proposed to manipulate surface waves dynamically. Literature \cite{reconfigEBG1} proposed tunable EBG structures based on varactors in early years. Recently, a reprogrammable topological insulator operating in microwave band is proposed \cite{reconfigEBG}. As Figure. \ref{fig:reconfigEBG} illustrates, configurations of elements are tuned by two states of PIN diodes, and the propagating direction of surface wave are modulated dynamically by switching states of elements. 

The high flexibility of R-MTS for manipulating surface wave shows it is a promising topic for future investigations.

\subsection{Nonlinear R-MTS}

Linear MTSs have experienced intensive research in past decades. To realize more functions, nonlinear effects need to be taken into consideration when designing MTSs. 
Strictly speaking, reconfiguration is a kind of nonlinearity, which occurs at the state-change moment. Nevertheless, at each stable working state, the response to incident EM wave is still linear. Therefore, the R-MTS is quasi-linear essentially.

For real nonlinear MTS, the response is always nonlinear. Generally, energy amplification, new frequency generation and some magnet-free nonreciprocity effects are certain forms of nonlinear effects. Based on these nonlinear effects, some nonlinear MTSs are designed over the years.

Energy amplification is based on the nonlinearity of materials. MTS amplifier based on transistors are implemented in early years \cite{1bitampreview,1bitampampmaintain,1bitamp2ampmaintain,1bitamp3ampmaintainref,2bitphaseampref}, and parametric amplification MTS element based on varactor is realized recently \cite{nonlinearthesis}.

\begin{figure}[!t]
	\centerline{\includegraphics[width=0.5\columnwidth]{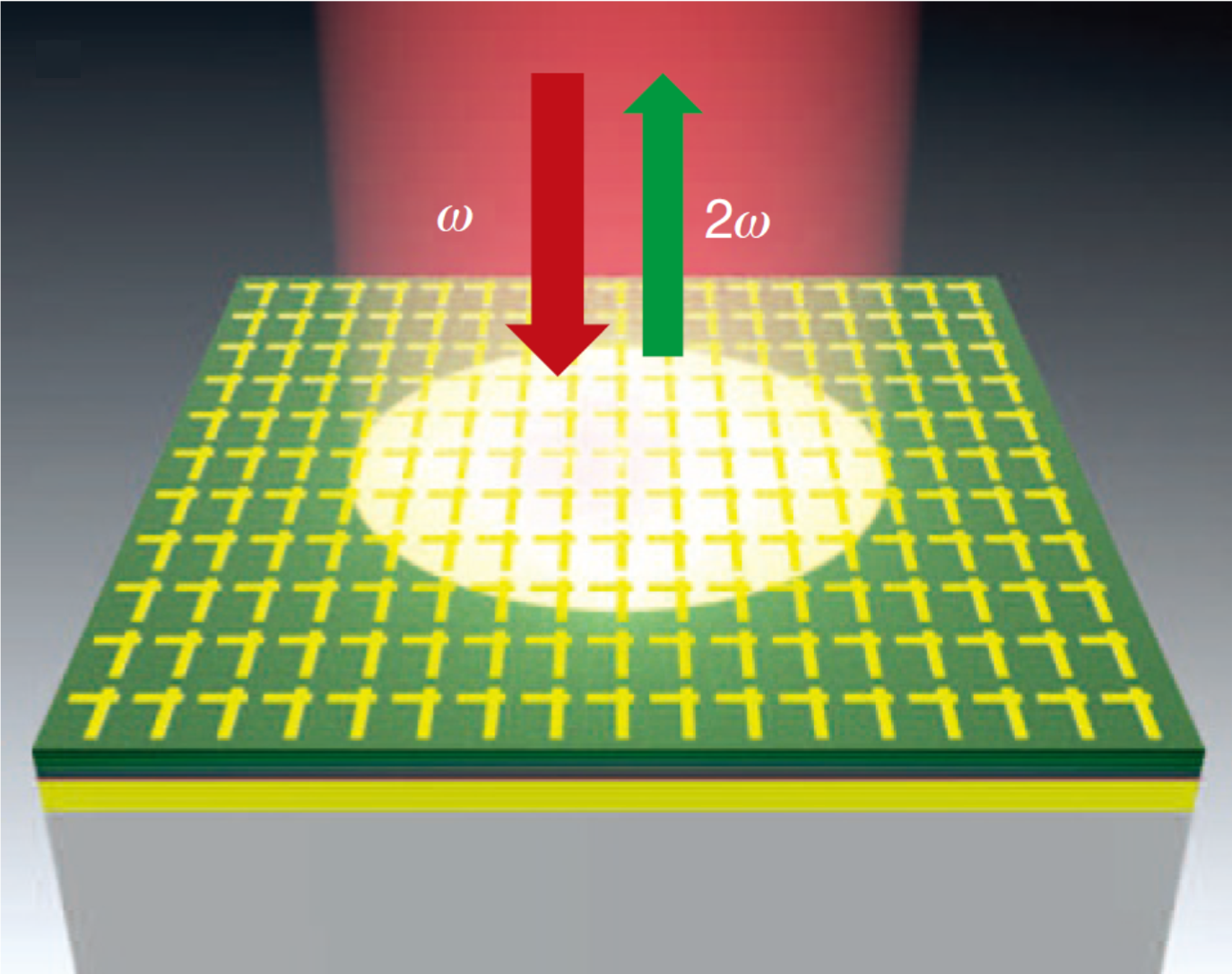}}
	\caption{Optical nonlinear MTS for second harmonic generation \cite{nonlinearoptical}.}
	\label{fig:opticalSHG}
\end{figure}

Frequency generation MTSs have been investigated since 1980s \cite{nonlinearsummary,nonlinearthesis}. Quasi-optical grid arrays are introduced with lumped nonlinear components to realize functions of spatial wave processing in microwave band, such as oscillator \cite{nonlinearoscillator,nonlinearoscillator1}, frequency doubler \cite{nonlineardoubler} and frequency mixer \cite{spatialmixer}. In optical frequency band, nonlinear materials are employed to form nonlinear MTS \cite{nonlinearoptical,nonlinearoptical3,nonlinearoptical1,nonlinearoptical4}, with second harmonic generation (SHG) or third harmonic generation (THG), as shown in Figure. \ref{fig:opticalSHG}. In recent years, with the development of time-modulated MTS, frequency reconfiguration using time dimension has become a new research trend \cite{freqtime2,freqtime3,freqtime4,freqtime5,2bitphasefreqFPGA,2bitphaseampfreqFPGA,2bitphaseampfreq3FPGA,2bitphasefreqdireFPGA,nbitampphasetime,2bitampfreq,nbitfreqpol,2bitpolfreq,2bitdirfreq}. 

Magnet-free nonreciprocal MTSs are emerging nowadays. To realize nonreciprocity without magnetic effects, nonlinear effects are taken into consideration. Based on the nonreciprocity of amplifier, nonreciprocal MTS can amplify the forward wave and block the backward wave \cite{nonreci,2bitdirectionmuldirection,nonreci1}. Besides, time-modulated R-MTS can also modulate frequency and beam direction without reciprocity \cite{timenonreci1,timenonreci2,timenonreci3}.

Apart from these applications, nonlinear MTS could also carry out other functions, such as all-optical logic gates \cite{swlogic}, spatial wave limiters \cite{swlimiter}, waveform-dependent absorbers \cite{sflimiter}  and so on, which are interesting topics for future research. Furthermore, reconfigurable nonlinear MTS can employ nonlinear effects with dynamic functions, which may lead a larger research tide in the future.

\section{Conclusion}
\label{secVII}

As the advanced form of MTS, R-MTS is proposed to manipulate the scattered wave dynamically. In recent years, numerous R-MTSs emerge with different manipulating dimensions and functions. In this review, we start with the interactions among R-MTS, EM wave and EM information in five dimensions, and propose the mathematical model of R-MTS in response to five dimensions of incident EM wave. Then we suggest a frame called information allocation strategy to categorize different types of R-MTSs systematically. Based on this strategy, 1-bit reconfigurable elements manipulating one of five dimensions are firstly reviewed and categorized. The advances of multi-dimensional and multi-functional 2-bit elements are reviewed and categorized in the following section. Various 2-bit elements are divided into two large categories, 2-bit-manipulating types (15 kinds) and multiplexed-manipulating types (25 kinds), and the existing ones are reviewed respectively in detail. 

Due to the rapid evolving speed of R-MTS, diverse terminologies appear, making the whole research architecture confusing. Hopefully, R-MTSs reorganized and reunited under information allocation strategy might provide a helpful viewpoint for researchers to find out the development thread and future research trends. In this paper, we find some types of the multi-bit reconfigurable elements are unrealized, like some 2-bit-manipulating elements and many multiplexed-manipulating elements, and the functions of R-MTS are not fully utilized. Besides, we have discussed some possible evolution clues of R-MTS such as multi-bit R-MTS, THz/optical R-MTS, surface-wave R-MTS, nonlinear R-MTS, and so on. 

As the recent researching highlight in both science and engineering fields, the exploration of R-MTS is just unfolding. R-MTS shows grand opportunities for critical applications in communication, detection, sensing, imaging and computing areas. It is believed that R-MTS will bring about a technological revolution in the near future. 

\acknowledgements{This work was supported in part by the National Natural Science Foundation of China under Grant U2141233, in part by ZTE Industry-Academia-Research Cooperation Funds, and in part by THE XPLORER PRIZE.}

\bibliographystyle{emsreport}
\bibliography{refs}

\end{document}